\documentclass[a4paper,fleqn,usenatbib,useAMS]{mnras}

\usepackage{array}
\usepackage{subfigure}
\usepackage{graphicx}	
\usepackage{amsmath}	
\usepackage{amssymb}	
\usepackage{multicol}        
\usepackage{bm}		
\usepackage{pdflscape}	

\usepackage{color}

\newcommand{\Mpc}{\ensuremath{\,{\rm Mpc}}}

\newcommand{\Hz}{\ensuremath{\, {\rm Hz}}}

\newcommand{\MHz}{\ensuremath{\, {\rm MHz}}}

\usepackage[normalem]{ulem}

\usepackage[T1]{fontenc}
\usepackage{ae,aecompl}

\usepackage{newtxtext,newtxmath}

\def\hi{\textsc{Hi~}}

\title[1/f noise]
{1/f Noise Analysis for FAST \hi Intensity Mapping Drift-Scan Experiment}

\author[Wenkai Hu et al.] 
{Wenkai Hu$^{1,5,7}$\thanks{Contact e-mail: \href{wkhu@nao.cas.cn}{wkhu@nao.cas.cn}},
Yichao Li$^{2}$,
Yougang Wang$^{1}$, 
Fengquan Wu$^{1}$,
Bo Zhang$^{3}$,
Ming Zhu$^{3}$,
\newauthor
Shifan Zuo$^{4}$,
Guilaine Lagache$^{5}$,
Yinzhe Ma$^{6,8}$,
Mario G. Santos$^{2,9}$,
Xuelei Chen$^{1,10,11,12}$\thanks{Contact e-mail: \href{xuelei@cosmology.bao.ac.cn}{xuelei@cosmology.bao.ac.cn}}
\\
$^{1}$ Key Laboratory of Computational Astrophysics, National Astronomical Observatories, Chinese Academy of Sciences, Beijing 100101, China\\ 
$^{2}$ Department of Physics and Astronomy, University of the Western Cape, Robert Sobukwe Road, Belville 7535, South Africa\\
$^{3}$ CAS Key Laboratory of FAST, National Astronomical Observatories, Chinese Academy of Sciences\\
$^{4}$ Department of Astronomy, Tsinghua University, Beijing 100084, China\\
$^{5}$ Aix Marseille Universit$\rm\acute{e}$, CNRS, LAM (Laboratoire d$^{\ \prime}$Astrophysique de Marseille), F-13388 Marseille, France\\ 
$^{6}$ School of Chemistry and Physics, University of KwaZulu-Natal, Westville Campus, Private Bag X54001, Durban 4000, South Africa \\
$^{7}$ ARC Centre of Excellence for All Sky Astrophysics in 3 Dimensions (ASTRO 3D), Australia\\
$^{8}$NAOC-UKZN Computational Astrophysics Centre (NUCAC), University of KwaZulu-Natal, Durban, 4000, South Africa \\
$^{9}$ South African Radio Astronomy Observatory (SARAO), 2 Fir Street, Observatory, Cape Town, 7925, South Africa\\
$^{10}$ School of Astronomy and Space Science, University of Chinese Academy of Sciences, Beijing 100049, China\\
$^{11}$ Department of Physics, College of Sciences, Northeastern University, Shenyang 110819, China\\
$^{12}$ Center of High Energy Physics, Peking University, Beijing 100871, China\\
} 

\pubyear{2021}

\begin{document}
\label{firstpage}
\pagerange{\pageref{firstpage}--\pageref{lastpage}}
\maketitle

\begin{abstract}
We investigate the 1/f noise of the Five-hundred-meter Aperture Spherical Telescope (FAST) receiver system using drift-scan data from an intensity mapping pilot survey. All the 19 beams have 1/f fluctuations with similar structures.  Both the temporal and the 2D power spectrum densities are estimated. The correlations directly seen in the time series data at low frequency $f$ are associated with the sky signal, perhaps due to a coupling between the foreground and the system response. We use Singular Value Decomposition (SVD) to subtract the foreground. By removing the strongest components, the measured 1/f noise power can be reduced significantly. With 20 modes subtraction, the knee frequency of the 1/f noise in a 10 MHz band is reduced to $1.8 \times 10^{-3}\Hz$,  well below the thermal noise over 500-seconds time scale. The 2D power spectra show that the 1/f-type variations are restricted to a small region in the time-frequency space and the correlations in frequency can be suppressed with SVD modes subtraction. The residual 1/f noise after the SVD mode subtraction is uncorrelated in frequency, and a simple noise diode frequency-independent calibration of the receiver gain at 8s interval does not affect the results. The 1/f noise can be important for HI intensity mapping, we estimate that the 1/f noise has a knee frequency $(f_{k}) \sim$ 6 $\times$ 10$^{-4}$Hz, and time and frequency correlation spectral indices $(\alpha) \sim 0.65$, $(\beta) \sim 0.8$ after the SVD subtraction of 30 modes. This can bias the \hi power spectrum measurement by 10 percent. 

\end{abstract}




\section{Introduction}
\label{sec:intro}
The 21cm line of the neutral hydrogen (\hi)
is an important probe for a wide range of astrophysical processes, such as star formation history \citep{1959ApJ...129..243S,1998ApJ...498..541K,2010AJ....140.1194B,2012ApJ...759....9K}, galaxy dynamics \citep{1999MNRAS.304..475M,2014A&A...566A..71L} and environmental dependence \citep{2010MNRAS.409..500O,2016MNRAS.457.4393J}, as well as tracing the cosmic large-scale structure \citep{2010ARA&A..48..127M,2012RPPh...75h6901P,2019arXiv190306212F,2020PASP..132f2001L}. 
Compared with optical survey, the 21cm line provides a good alternative way of tracing large-scale structure in the radio wavelength. A number of HI surveys have been carried out, e.g. the HIPASS survey \citep{2004MNRAS.350.1195M,2004MNRAS.350.1210Z}, and the ALFALFA survey \citep{2005AJ....130.2598G,2007AJ....133.2087S,2007AJ....133.2569G}, and JVLA deep survey \citep{2014arXiv1401.4018J}. However, limited by the  sensitivity of the telescopes, the redshift range of these surveys are much smaller than the current optical surveys.

It has also been proposed that the large scale structure can be more efficiently mapped at low angular resolution by applying the  {\it intensity mapping} (IM) method \citep{2008PhRvL.100i1303C}. A number of such experiments are underway, such as  
Tianlai \citep{2012IJMPS..12..256C,2015ApJ...798...40X,Li:2020ast,wu2020tianlai}, CHIME \citep{2014SPIE.9145E..22B,2014SPIE.9145E..4VN} and
HIRAX \citep{2016SPIE.9906E..5XN}, as well as the specially designed single dish experiment BINGO \citep{2012arXiv1209.1041B,2016arXiv161006826B}. This technique
may also be applied to the new generation of general purpose radio telescopes, such as the MeerKAT \citep{2017arXiv170906099S} and the Square Kilometer Array (SKA) \citep{2015aska.confE..19S} in the southern hemisphere, and the Five-hundred-meter Aperture Spherical Telescope (FAST) \citep{2011IJMPD..20..989N} in the northern hemisphere. For forecasts of the FAST \hi intensity mapping survey, see e.g. \citet{2016ASPC..502...41B} and \citet{Hu:2019okh}.

In a radio observation, the time stream data $d(t, \nu)$ can be modeled as 
\begin{eqnarray}
d(t, \nu) = T_{\rm in}(t, \nu) G(t, \nu) + n(t, \nu),
\end{eqnarray}
where $T_{\rm in}(t, \nu)$ is the input temperature, $G(t, \nu)$ is the power gain of the system, and $n(t, \nu)$ is the receiver noise.  If we use an overbar to define the time averaged quantities, to first order, the varying part $\delta_d(t, \nu) \equiv d(t, \nu)/\bar{d}(\nu) - 1$ is given by
\begin{eqnarray}
\delta_d(t, \nu) &\approx& \frac{\delta T_{\rm in}(t, \nu)}{\bar{T}_{\rm in}(\nu)}  + \frac{\delta G(t, \nu)}{\bar{G}(\nu)} + \frac{n(t,\nu)}{\bar{T}_{\rm in}(\nu) \bar{G}(\nu)}.
\end{eqnarray} 
The input signal, the variations in the gain and the noise all give rise to  variations of the data.  

It has long been noted that for electronic devices, besides the thermal noise which in the range of interest behaves as white noise, there is also a time-correlated noise, often referred to as the 1/f noise or flicker noise, which has also been found to be present in the spectrum of a wide variety of systems (for a brief review of its research history, see e.g. \citealt{2002physics...4033M}). In many cases involving electronic circuits, the 1/f noise may originate from the fluctuations of the amplifier gain. Such fluctuations leave long-range correlations in time, and in sky survey it may contaminate the final intensity map. For Cosmic Microwave Background (CMB) experiments \citep{1996astro.ph..2009J}, several methods have been explored to suppress the 1/f noise \citep{2002A&A...387..356M,2002A&A...391.1185S,2004A&A...428..287K,2009A&A...506.1511K,2010MNRAS.407.1387S}. For \hi IM survey, the effect of 1/f noise has been analyzed through simulations \citep{2015MNRAS.454.3240B,2018MNRAS.478.2416H,Chen:2019jms}. 
\citet{2015MNRAS.454.3240B} presented simulations of single-dish observations including an instrument noise model with 1/f and white noise, and sky emission with a diffuse Galactic foreground and \hi emission. They found that using the principal component analysis (PCA), one can remove 1/f noise contamination down to the thermal noise level.
In \citet{2018MNRAS.478.2416H}, the power spectral density of the gain fluctuations are modeled as a power law, and characterized by parameters $\alpha$ and $\beta$ (will be defined in Eqs.\ref{psd} and Eqs.\ref{psd2d}). The degree of 1/f noise frequency correlation was found to be critical to the success of \hi IM experiments. Using the current component separation techniques, the removal is easier for the case with small value of $\beta$ ($\beta < 0.25$). In \citet{Chen:2019jms}, the impact of the 1/f noise on SKA cosmological parameter measurement is forecasted with the Fisher matrix formalism.

Recently, \citet{2020arXiv200701767L} developed an 1/f noise power spectrum density parametrization, and measured the power spectrum density of the 1/f noise for the MeerKAT receiver system using the data of a tracking observation. They applied the Singular Value Decomposition (SVD) to the data set 
to remove the external temperature fluctuations from the sky variations and recovered the system induced 1/f noise power spectrum density.
The result show that the system induced 1/f-type variations are well under the thermal noise fluctuations over the time scale of a few hundred seconds.

Here we measure the 1/f noise using the data collected in the FAST drift-scan intensity mapping experiment.
The layout of this paper is as follows. In Sec.~\ref{sec:data} we describe our observation data. In Sec.~\ref{sec:processing}, we present our data processing pipeline, including flagging, bandpass normalization, singular value decomposition (SVD) and gap filling; we also introduce the model for describing the 1/f noise, and measure the temporal and 2-D power spectrum density of FAST 1/f noise. In Sec.~\ref{sec:Results}, we present the results and quantify the 1/f noise influences on \hi power spectrum by simulations. In Sec.~\ref{sec:Discussions} we discuss the results. Finally we summarize the results in Sec.~\ref{sec:Summary}.

\begin{figure}
    \centering
    \includegraphics[width=0.49\textwidth]{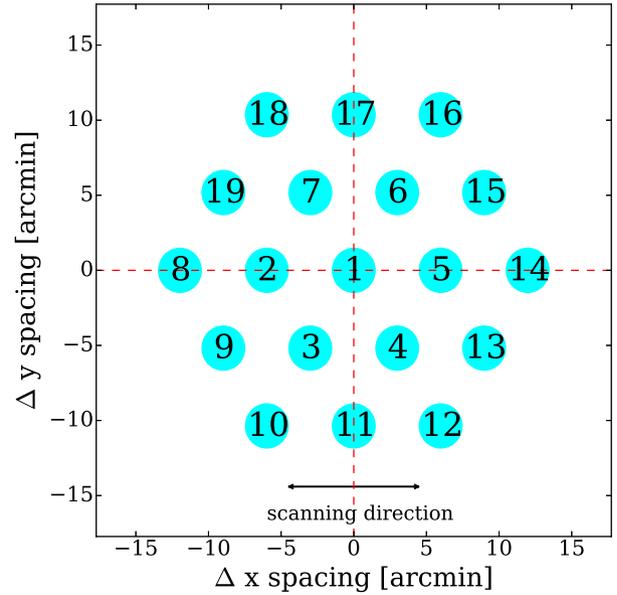}
    \caption{The beams of the FAST L-band 19-beam receiver (see \citealt{2020RAA....20...64J} for details). The beams are arranged in two concentric hexagonal rings around the central beam. The beams are labeled with numbers. The colored area shows the FWHM (Full Width at Half Maximum) of the beam at redshift zero for HI observation ($\sim$ 3 arcmin). The drift scan direction is paralleled to the symmetry axis of the hexagon.}
    \label{feeds}
\end{figure}

\section{Data}
\label{sec:data}

The FAST has a very large aperture (300 meters during operation) and is equipped with multiple feeds and receivers, ideal for conducting large surveys.
Our observations were carried out with the L-band 19-beam feed system and the associated cryogenic receiver. The feeds are arranged in two concentric hexagonal rings around the central one \citep{2018IMMag..19..112L}. Figure~\ref{feeds} shows the FAST L-band Array of 19 feed-horns. The feeds are labeled with numbers for clarity.

In this work we use data collected by the FAST \hi intensity mapping pilot experiment during the risk-shared observations made in May 2019. Five strips spanning $125^{\circ} < {\rm R.A.} < 325^{\circ}$ and $25.4^{\circ} < {\rm Dec.} < 26.7^{\circ}$ are drift scanned in five consecutive days (27 May to 31 May).  Each strip is observed for 13.3 hours. The frequency ranges from $1.0$ GHz to $1.5$ GHz. We shall use the third-day (29 May) data for most of our analysis, though we have checked that the other days give similar results. This data set has a time resolution  of 0.1 s, and a frequency resolution of 7.63 kHz. Considering our scientific objectives and computing capacity, we further re-bin the data into time and frequency resolution of 1.0 s and 0.2 MHz. For the calibration, the built-in noise diode of the feed is fired on for 1 s in every 8 s. During this pilot survey, we have tried both low (1 K) and high (10 K) setting for the noise diode effective temperature, though for the data analyzed here (collected 29 May) it is the former case. For comparison, the system temperature is about 20 K. We use the data when the noise diode is off to measure the 1/f noise, and fill in the gaps through interpolation (see Sec. \ref{sec:processing}).

\begin{figure}
    \centering
    \includegraphics[width=0.48\textwidth]{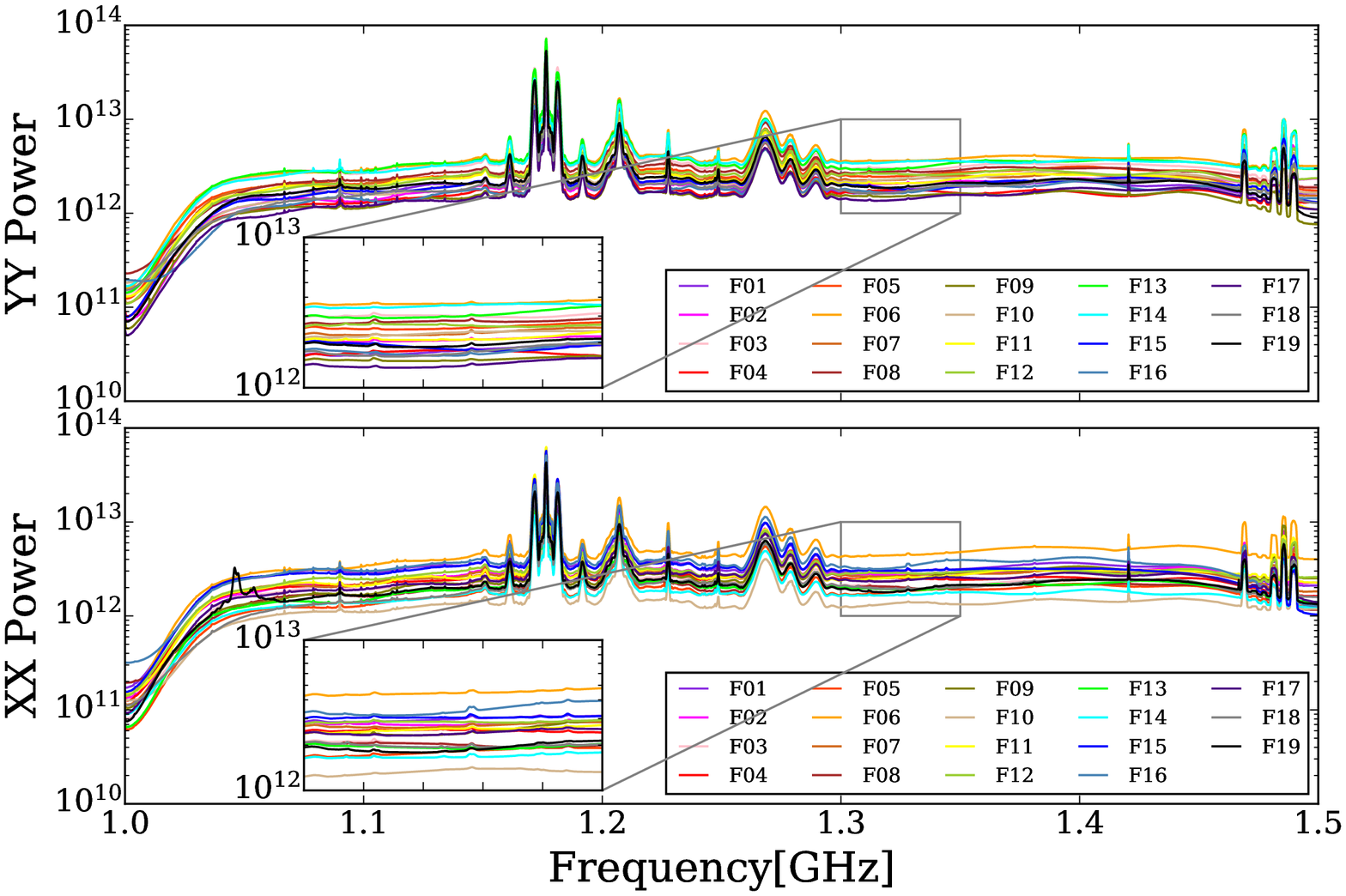}
    \includegraphics[width=0.48\textwidth]{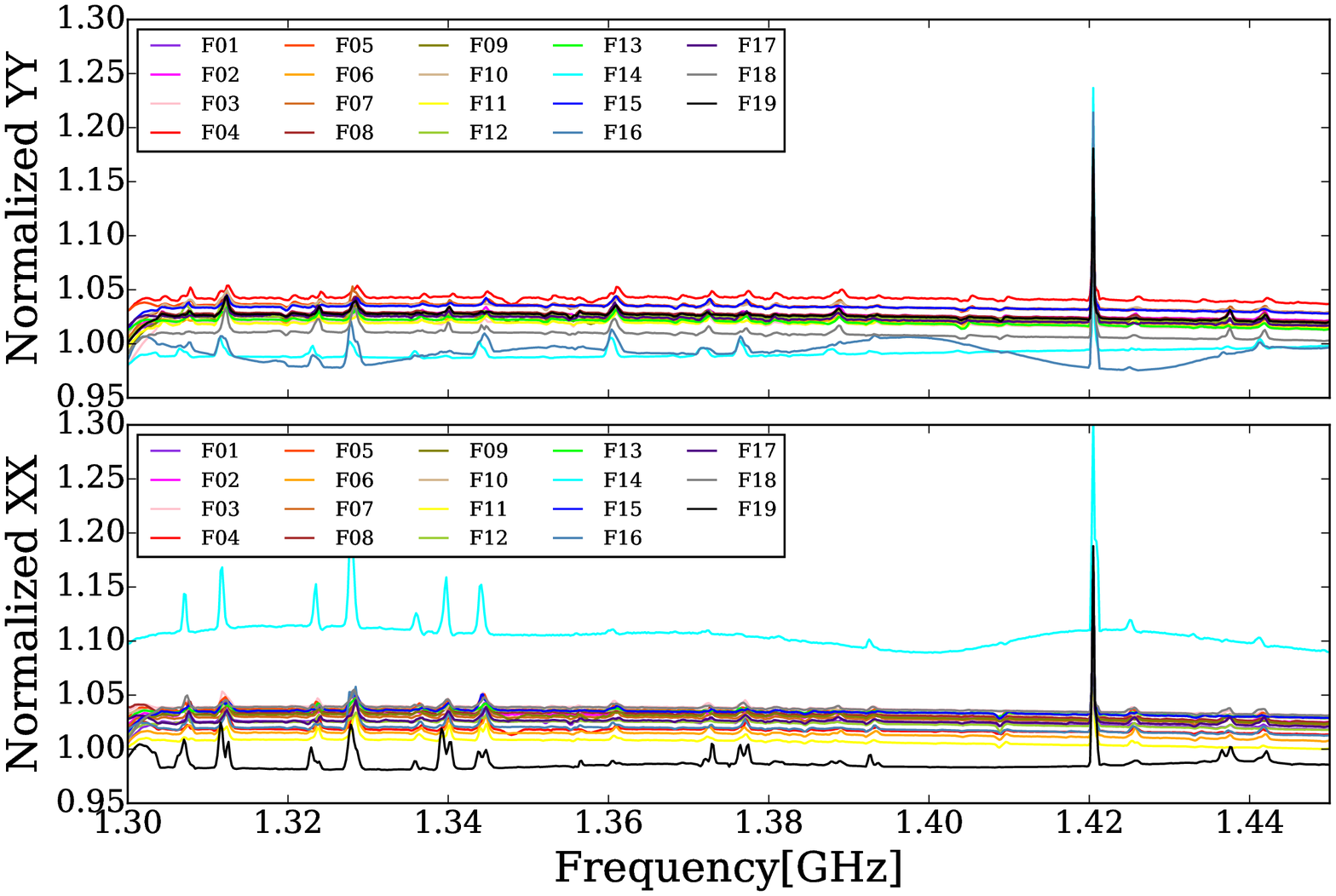}
    \caption{The raw (upper panel) and bandpass normalized (lower panel) frequency spectra for the 19 feeds of the FAST L-band multi-beam receiver feeds. The raw spectra are obtained by taking the time average of the raw data (in ADC units) through the whole time of observation. The bandpass normalized spectra are the medians of the spectra which are normalized by the time-averaged data.
    The two polarizations are shown in the top and bottom subpanel.
    In order to show the RFI more clearly, a 1.3GHz to 1.35GHz zoom-in sub figure is also shown in each panel.}
    \label{freq_spec}
\end{figure}

Figure~\ref{freq_spec} shows the raw (upper panel) and bandpass normalized (lower panel) frequency spectra of the FAST data for the 19 beams, each with one curve. The bandpass shape is normalized by dividing the time-averaged data.
The Radio Frequency Interference (RFI) dominates the frequency range from 1.15 GHz to 1.3 GHz, as well as the the range beyond 1.45 GHz. In the range of 1.05 GHz to 1.15 GHz, there are also many smaller peaks due to chronically present RFI. A 1.3 GHz to 1.35 GHz zoom-in subfigure is also shown, where the small peaks are also due to RFI. The receiver response drops at the two ends of the frequency band. Considering the data quality, we only use the data in the frequency range of 1.3 GHz to 1.45 GHz, which is relatively clean. The bandpass normalized spectra show there are some chronically present RFI, which will all be masked in our pipeline. The peaks at 1420 MHz are from the Galactic neutral hydrogen.

\begin{figure}
    \centering
    \includegraphics[width=0.48\textwidth]{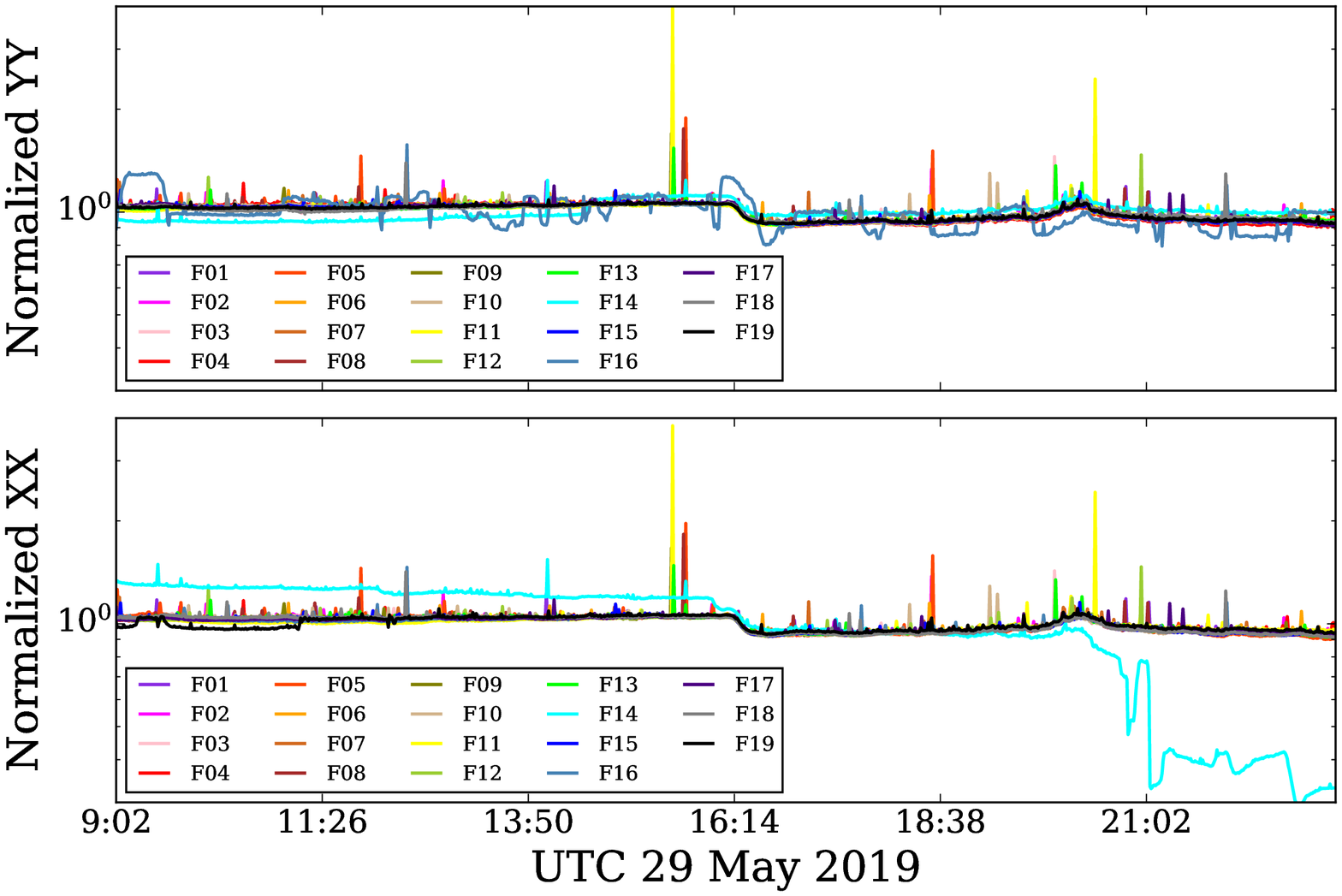}\\
    \includegraphics[width=0.48\textwidth]{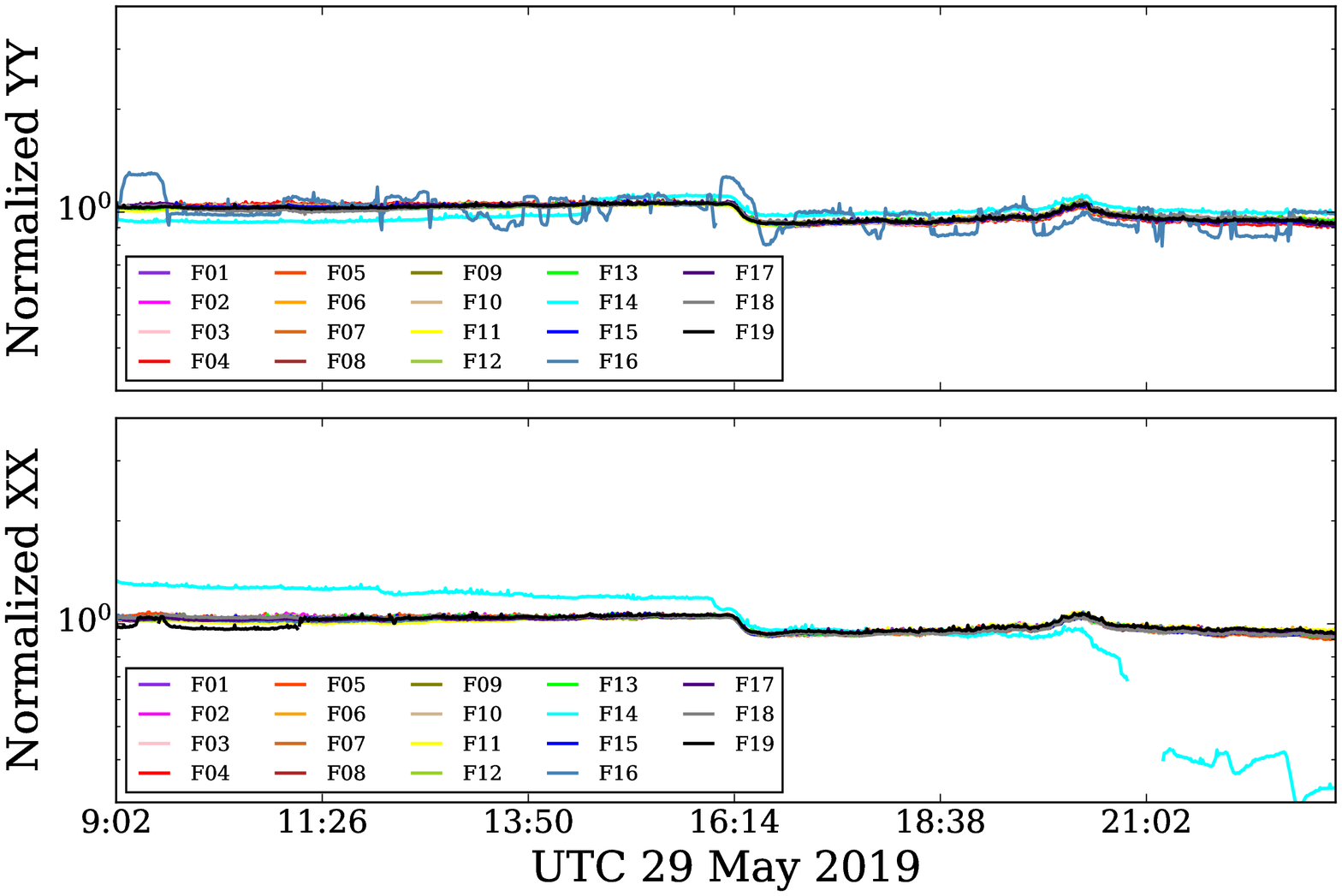}
    \caption{ Top:
    The frequency-averaged time stream data (bandpass normalized) of the FAST 19 L-band feeds. The observation on 29th May starts from 09:02 and ends at 23:14. These time spectra are obtained by taking the average of the data across frequencies from
    $1.3$ GHz to $1.45$ GHz. Bottom: Same as above, but after removing the flagged data.} 
    \label{time_spec}
\end{figure}

Figure~\ref{time_spec} shows the bandpass normalized time stream data for each of the 19 feeds. The top panel shows the time stream data obtained by taking the average of the bandpass normalized power across frequencies from 1.3 GHz to 1.45 GHz.  The spikes along the time stream data are caused by the transit of celestial radio sources.
A few Channels (the YY polarization of Feed 16 and the XX polarization of Feed 14) 
are unstable during this observation.
Except for these, the mean power of each feed varies slowly during the observation. These variations appear to be nearly synchronous for different beams, they are probably produced by the sky mean temperature variations.

Besides the variations of sky temperature, we also found the overall amplitude has some variations during the observing run. Figure~\ref{time_spec} shows a jump of the signal amplitude at about 16:14 UTC (near midnight local time) on 29th May 2019 for all beams. We also find a similar jump in the observing run on 28th May 2019, but not on 30th, May 2019. The jumps we found all happened at late night, but not exactly the same time, and the whole jump took 20 to 30 minutes. As we learned, similar jumps were also found in some other observing runs of FAST. At present, it is not known what caused such jumps, but these jumps do not affect our result much, and such variations can be removed together with the sky variations by subtracting the first SVD mode.

\section{Data Processing}
\label{sec:processing}
In the above we looked at the general behavior of the raw data, next we process the data for quantitative analysis. The steps of the data processing include RFI flagging, bandpass normalization, sky variations subtraction and RFI gap filling.

\subsection{RFI Flagging and Bandpass Normalization}

The RFI flagging is performed along the time axis first. We take the frequency averaged data, evaluate the root mean square (r.m.s., $\sigma$) along the time axis, and mask the data points with value greater than $5.5 \sigma$ as RFI-contaminated. The data with time stamps within 10 steps (i.e. 10 seconds) around such RFI points are also masked.
Because the variations of the band baseline across time may affect the r.m.s estimation, we fit the band baseline during the flagging.
The band baseline is evaluated by averaging the data across frequencies and then smoothed with a low-pass filter, 
as realized with the {\tt filtfilt} function \citep{van1995python,2020SciPy-NMeth}. The spikes of some strong point sources could also be masked as RFI in this processing, but for our purpose here it does not matter, as we are studying the noise, not the source. We iterate the flagging process, so that weaker RFI points can be detected with better sensitivity after the stronger ones are removed. The process is repeated until there is no more RFI point flagged.

We then perform the flagging along the frequency axis. We first remove a few frequency channels which are known to be contaminated by some chronically present RFI.  As shown later, such 
frequently-appeared RFI points induce extra correlations across frequencies.
We then estimate the r.m.s. of the time
averaged data and mask the frequencies with values greater then 
$5.5 \sigma$. The frequencies within 1 MHz to these RFI frequencies are also masked in case they are contaminated. A total of 17 RFI points in frequency are masked, which occupies 28\% of the frequencies in the band considered here. The RFI-flagged time-stream data for all 19 feeds is shown in the bottom of Figure~\ref{time_spec}.

As here we are interested in the fluctuations around the time average, 
the absolute flux calibration is not necessary for our analysis. 
However, the relative bandpass shape needs to be calibrated. 
We normalize the flagged data with the time average:
\begin{eqnarray}
    d^{c}(t,\nu) = \frac{d(t,\nu)}{\langle d(t,\nu)\rangle_{t}}.
    \label{bandpass_cali}
\end{eqnarray}
As the strong point sources have been flagged, this time average is given by the galactic continuum emission plus unresolved sources in the surveyed part of the sky. The normalization should calibrate out the system bandpass gain and mean spectrum features from the sky.

Before we calculate the power spectrum, we also need to fill in the masked regions and noise-diode-on regions, otherwise they will introduce unwanted structures in the power spectrum. We first apply a low-pass filter to the data in order to obtain the band baseline, and then interpolate the band baseline along the frequency direction using cubic spline. Finally we fill in the masked region with Gaussian white noise, with $\sigma$ given by the standard deviation of the unmasked region of the flagged data and the mean value set by the baseline value obtained by interpolation. We then use similar interpolation method to fill in the gaps along the time direction. As this affects only a small fraction of the whole data set, it is unlikely to change the result significantly.

\subsection{Power Spectrum}

The power spectrum density is estimated via Fourier transform, the 
algorithm used is similar to that described in \citet{2020arXiv200701767L}.  
The temporal power spectrum density function, $\hat{S}^{t}(f,\nu)$ is estimated by
\begin{eqnarray}
\hat{S}^{t}(f,\nu) &=&\bigg\lvert \sqrt{\frac{\delta t}{n_{t}}}\sum_{n=1}^{n_{t}}d_{t}(\nu)\exp[-2\pi if n\delta t]\bigg\rvert^{2}\\
&=&\frac{\sigma_{\rm in}^2}{\bar{T}_{\rm in}^2}(1 + S^{t}(f,\nu)),
    \label{fft1d}
\end{eqnarray}
where $f$ is the temporal frequency ($f$ = 1/t), $n_{t}$ is the number of time samples,
$\sigma_{\rm in}^2 = T_{\rm in}^2/\delta\nu$. If we ignore the correlations between the different frequency channels of the electromagnetic wave,
the temporal power spectrum density function can be modeled as:
\begin{eqnarray}
S^{t}(f) = \frac{A}{\delta\nu}(1 + (f_{k}/f)^{\alpha}),
    \label{psd}
\end{eqnarray}
where $A = (T_{\rm in}/\bar{T}_{\rm in})^2$, $\alpha$ is the spectral index, and
$f_{k}$ is the knee frequency, defined as the point where 1/f noise
has equal power as the white noise. Note as shown in Eq.~(\ref{psd}), 
$f_{k}$ depends on electromagnetic wave frequency resolution, because the white noise power is inversely proportional to the square root of the frequency resolution, 
$\sim 1/\sqrt{\delta \nu}$. 

Taking the electromagnetic wave frequency correlation into consideration, the 2-D power spectrum density can be estimated by Fourier transforming the observed time stream data along both time and frequency axes.
We adopt a simple power-law model:
\begin{eqnarray}
\hat{S}(f,\tau) =\bigg\lvert \sqrt{\frac{\delta t\delta\nu}{n_{t}n_{\nu}}}\sum_{n=1}^{n_{t}}\sum_{m=1}^{n_{\nu}}\exp[-2\pi i(fn\delta t + \tau m\delta\nu)]\bigg\rvert^{2},
    \label{fft2d}
\end{eqnarray}
where $n_{\nu}$ is the number of electromagnetic wave frequency samples. Including frequency correlation, the 2-D power spectrum density can be expressed as:
\begin{eqnarray}
S(f,\tau) = A\bigg(1 + \frac{1}{K\delta \nu}
\bigg(\frac{f_{k}}{f}\bigg)^{\alpha}\bigg(\frac{\tau_{0}}{\tau}\bigg)^{\frac{1-\beta}{\beta}}\bigg),
    \label{psd2d}
\end{eqnarray}
where $\tau= 1/\nu$ is the spectroscopic frequency, and 
$\tau_0=1/(n_\nu \delta \nu)$.
The spectral index of the electromagnetic wave frequency correlation is defined by $\beta$ with 
$0 < \beta < 1$. When $\beta=1$, 
$S(f,\tau)$ is independent of frequency, and 
$\beta=0$ means that the 1/f noise is fully correlated in every frequency channel.
$$K=\int {\rm d}\tau ~{\rm sinc}^2(\pi\delta\nu\tau)
\left(\frac{\tau_0}{\tau}\right)^{(1-\beta)/\beta}$$ 
represents the 
relation of 1/f noise power and frequency resolution. 

\subsection{Sky Variations Subtraction}
Besides system noise, the variation of the sky signal in the drift scan 
also contributes to the observed 1/f power spectrum. This comes from the
temperature variation of the continuum emission along the drift-scan path, and it varies synchronously across frequencies.

We apply the Singular Value Decomposition (SVD) to the bandpass normalized data $d^{\rm c}$ (Eq.~(\ref{bandpass_cali}))
and subtract the sky variations by removing the first few principal components to obtain the cleaned data $d^{\rm clean}$:
\begin{eqnarray}
d^{\rm c} = U \Lambda V^T,\,\,
d^{\rm clean} = \left( I - \sum_{i}u_{i}^{}u_{i}^T \right)d^{\rm c},
\end{eqnarray}
in which the superscript $^T$ indicates the transpose of the matrix;
$d^{\rm c}$ is the bandpass normalized time stream data with array shape of $n_{\nu}\times n_{t}$; the columns of 
$U=\{u_0, u_1, ..., u_{n_{\nu}}\}$ and $V=\{v_0, v_1, ..., v_{n_{t}}\}$ are the spectroscopic and temporal modes, respectively.

\begin{figure}
    \centering
    \includegraphics[width=0.47\textwidth]{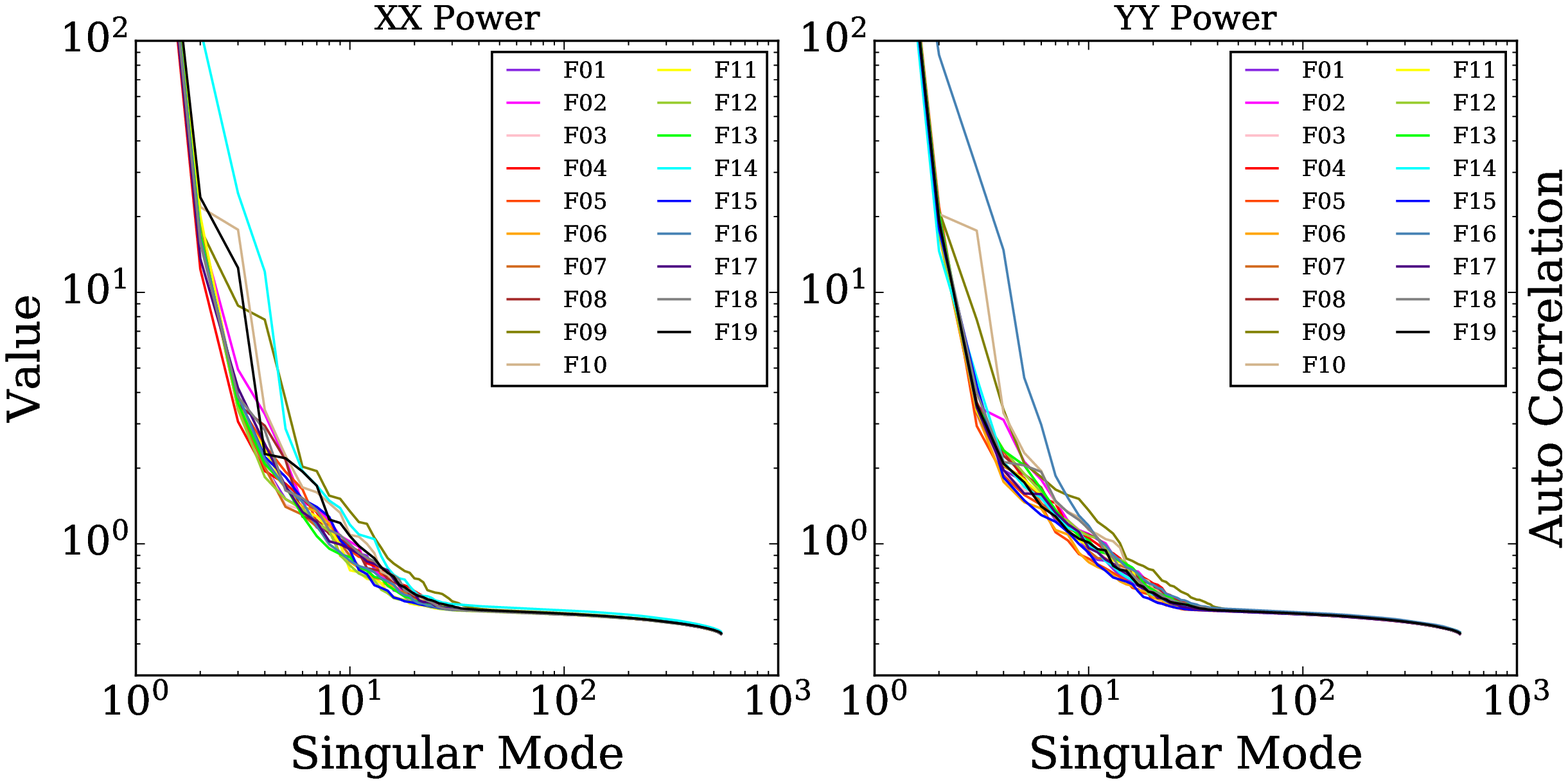}
    \includegraphics[width=0.47\textwidth]{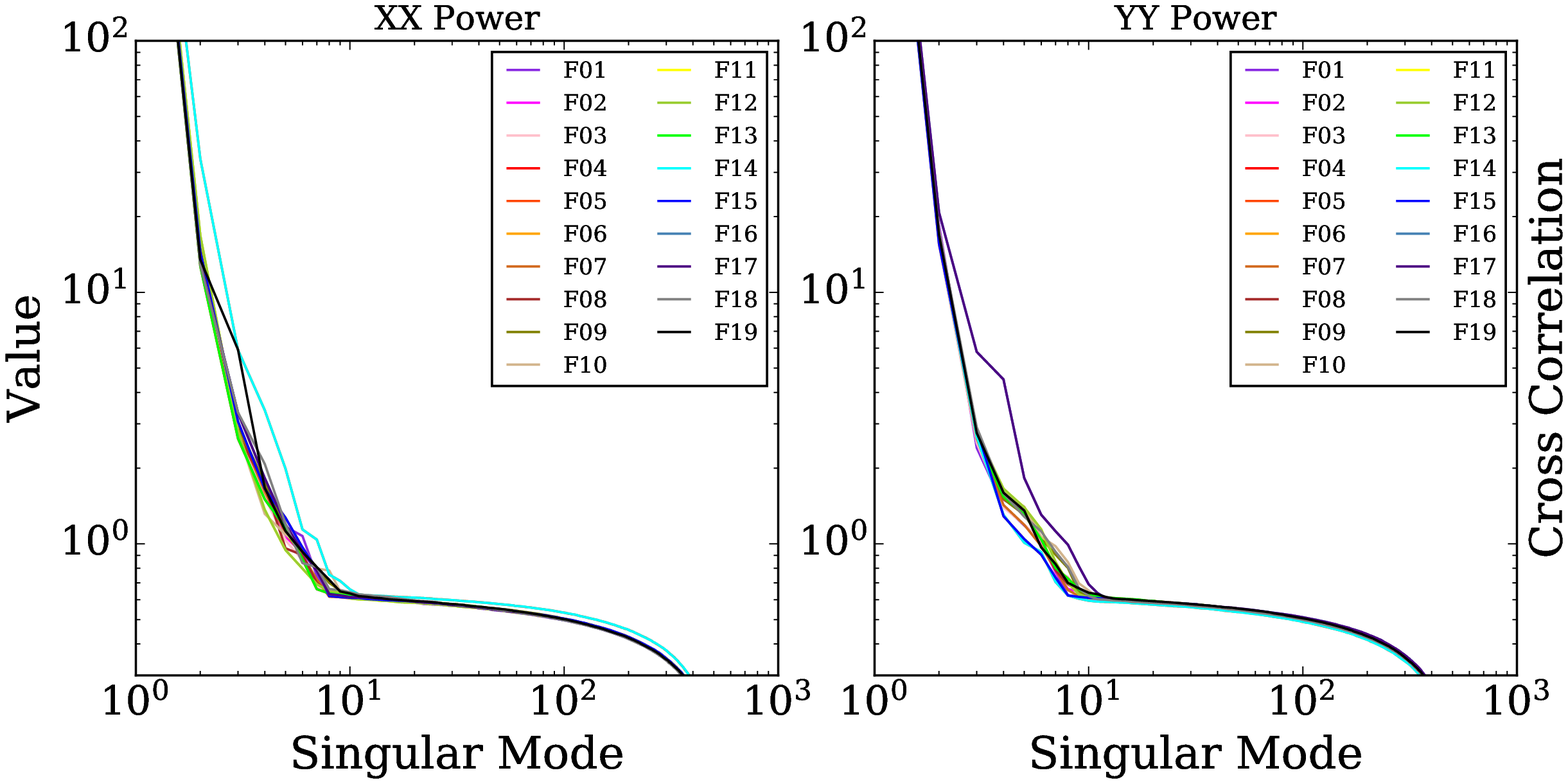}
    \caption{
    The singular values of auto-correlation (top panels) and cross-correlation 
  (bottom panels) for each of 19 feeds. The tow polarizations are shown 
    in the left and right panels. For the cross-correlation result of each feed, we only show the singular values of the cross-correlation with its nearest neighbouring feed in the same horizontal line along scanning direction as an example. To compare with auto-correlation SVD, the singular values of cross-correlation SVD are the square root of original values, $\sqrt{{\rm diag}\{\Lambda_\times\}}$.}
    \label{Smode}
\end{figure}

We also consider a sky subtraction method that uses the SVD modes estimated from the cross correlation of the data from two feeds observing the same position of the sky. We use the frequency-frequency covariance matrix for two feeds,
\begin{eqnarray}\label{autosvd}
C_{\nu\nu^{\prime}} = \langle d^{\rm c}_{A} d^{\rm c^T}_{B}\rangle = U_{A}^{}\Lambda_{\times}^{} U_{B}^T,
\label{covariance}
\end{eqnarray}
in which, 
$d^{\rm c}_{A}$ and $d^{\rm c}_{B}$ are the time stream data of feeds $A$ and $B$; and the columns of $U_A$ and $U_B$ are the line-of-sight (LoS) modes of $d^{\rm c}_{A}$ and $d^{\rm c}_{B}$ respectively,
\begin{eqnarray}
U_A = (a_{1}, a_{2}, ... , a_{n_{\nu}}),\,\,  U_B = (b_{1}, b_{2}, ... , b_{n_{\nu}}),
\label{SVD_modes}
\end{eqnarray}
where $n_{\nu}$ is the number of frequency points of the data. 
Finally the cleaned data with $N$ LoS modes removed can be obtained by:
\begin{eqnarray}
d_{A}^{\rm clean} = \left( I - \sum_{i=1}^{N}a_{i}^{}a_{i}^T \right)d^{\rm c}_{A},
\label{cleaned_map}
\end{eqnarray}
where $I$ is the identity matrix, $i$ refers to the $i$-th LoS mode. 
For identical $A, B$, the cross-correlation is reduced to the auto-correlation case. In the data processing for a feed, we take its bandpass normalized time stream data as $d^{\rm c}_{A}$ and cross correlate it with the other feeds in the same line along the scanning direction.

We apply the SVD mode subtraction to the bandpass normalized time steam data of each feed and polarization. The singular values for the auto-correlation data of each feed are shown in the top panels of Figure~\ref{Smode} in rank order. We can see the first few singular values are particularly large, but the singular values drop quickly, by more than an order of magnitude on average for the first $30$ modes subtracted. The singular values eventually reach a flat floor, which indicates the thermal noise level of the time stream data. 

\begin{figure}
    \centering
    \includegraphics[width=0.48\textwidth]{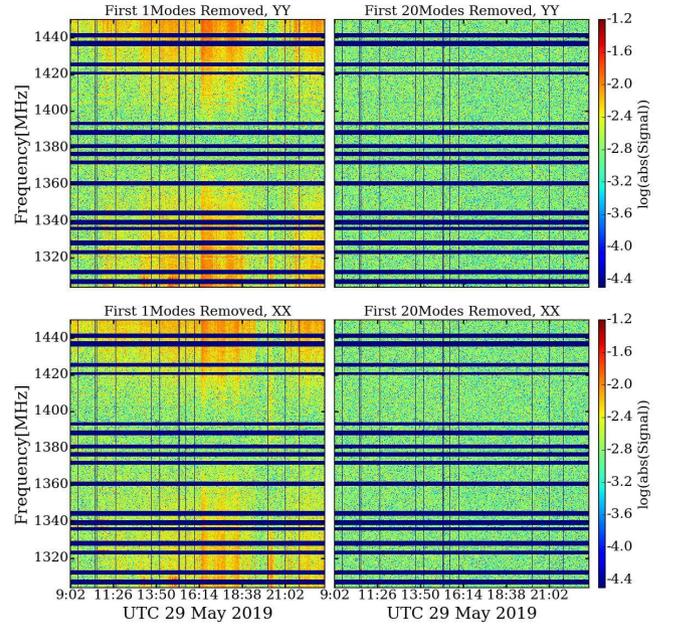}
    \caption{The waterfall plots of the bandpass normalized data after 1 mode (left panels) or 20 modes (right panels) removed for Feed 1. The upper row shows the YY polarization and lower row shows the XX polarization.}
    \label{map_svd}
\end{figure}

The processed (flagged and bandpass normalized) time stream data (Feed 1 as an example) with different number of SVD mode subtraction are shown with the waterfall plot in Figure~\ref{map_svd}. The deep blue lines and strips are the masked regions. 
The obvious structures of fluctuations are removed with the subtraction of the major SVD modes. 

\begin{figure}
    \centering
    \includegraphics[width=0.48\textwidth]{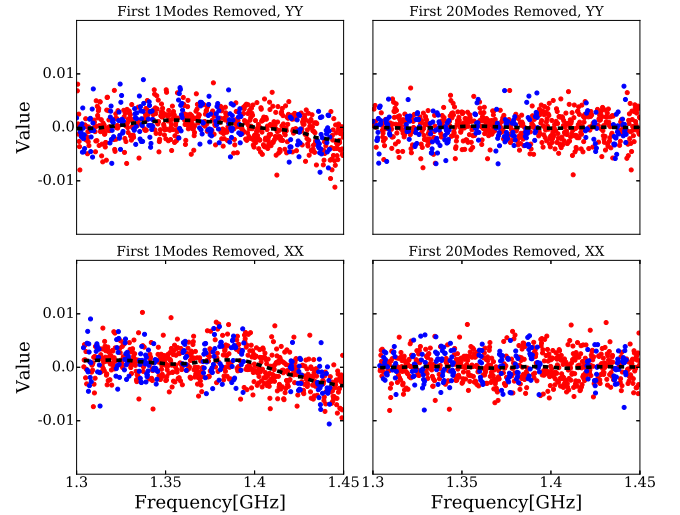}
    \caption{An example of the filled frequency spectra at a given time point, for first 1 and 20 SVD modes removed. The upper and lower row refer to the YY and XX polarization, respectively. The masked regions (blue points) are filled with Gaussian white noise. The black dotted lines show the fitted baseline.}
    \label{filling}
\end{figure}

After removing the singular modes,  we fill in the masked regions and the  noise-diode-on regions. With the first few SVD modes removed, the wide-band structures disappear and the average spectrum is nearly flat and close to zero. After filling for frequency spectrum at every time point, we do the same filling in time spectrum at each frequency point. Finally, the data is ready for power spectrum analysis. Figure~\ref{filling} shows an example of filled frequency spectra for data at a given time point with the first 1 and 20 SVD modes removed.

\begin{figure*}
    \centering
    \includegraphics[width=0.96\textwidth]{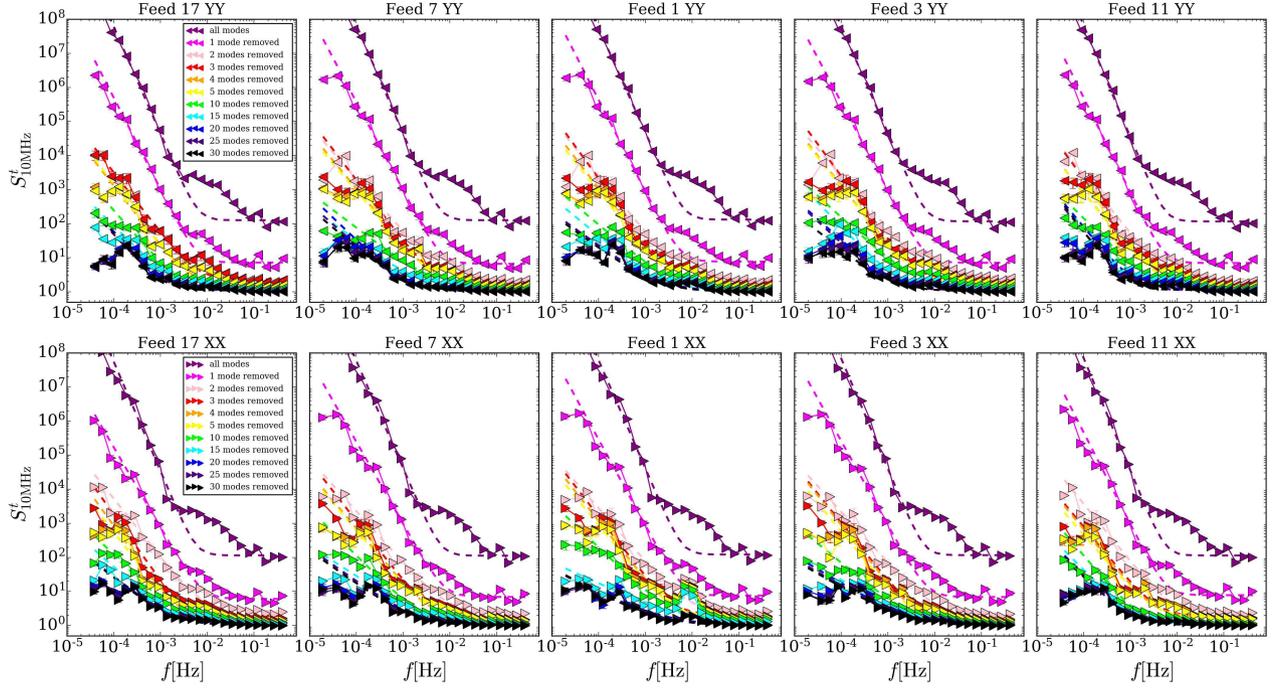}
    \caption{
    The frequency averaged temporal power spectra of time stream data cleaned with cross-correlated SVD modes.
    The top and bottom subpanels are the YY and XX polarization;
    From left to the right subpanels, the results are for the cross correlations of
    Feed 17, Feed 7, Feed 1, Feed 3 and Feed 11 with its nearest neighbouring feed along scanning direction, respectively.
    The dashed lines show the best-fit temporal power spectrum model of Eq. ~(\ref{psd}). 
    The thermal noise part of these spectra are normalized to $1$.}
    \label{tps_cov}
\end{figure*}

\begin{figure*}
    \centering
    \includegraphics[width=0.96\textwidth]{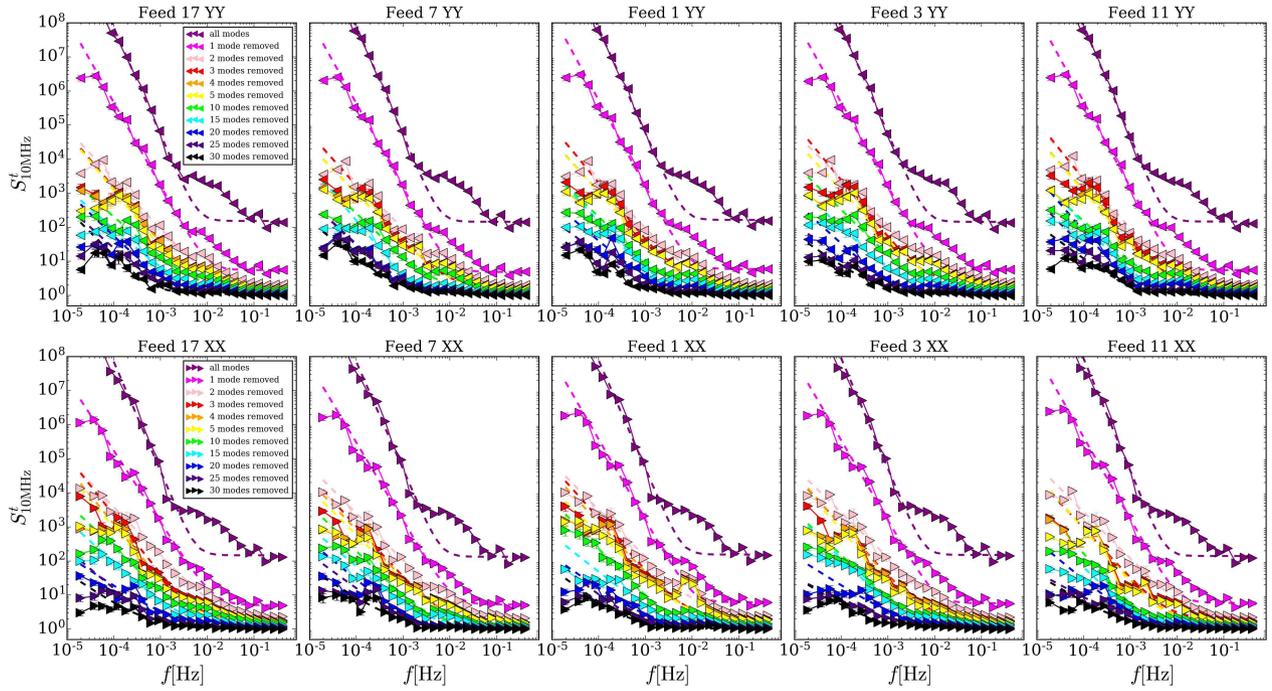}
    \caption{
    Same as Figure~\ref{tps_cov}, but using time stream data cleaned 
    with auto-correlated SVD modes.}
    \label{tps}
\end{figure*}

\begin{figure*}
\begin{multicols}{2}
\includegraphics[width=0.48\textwidth]{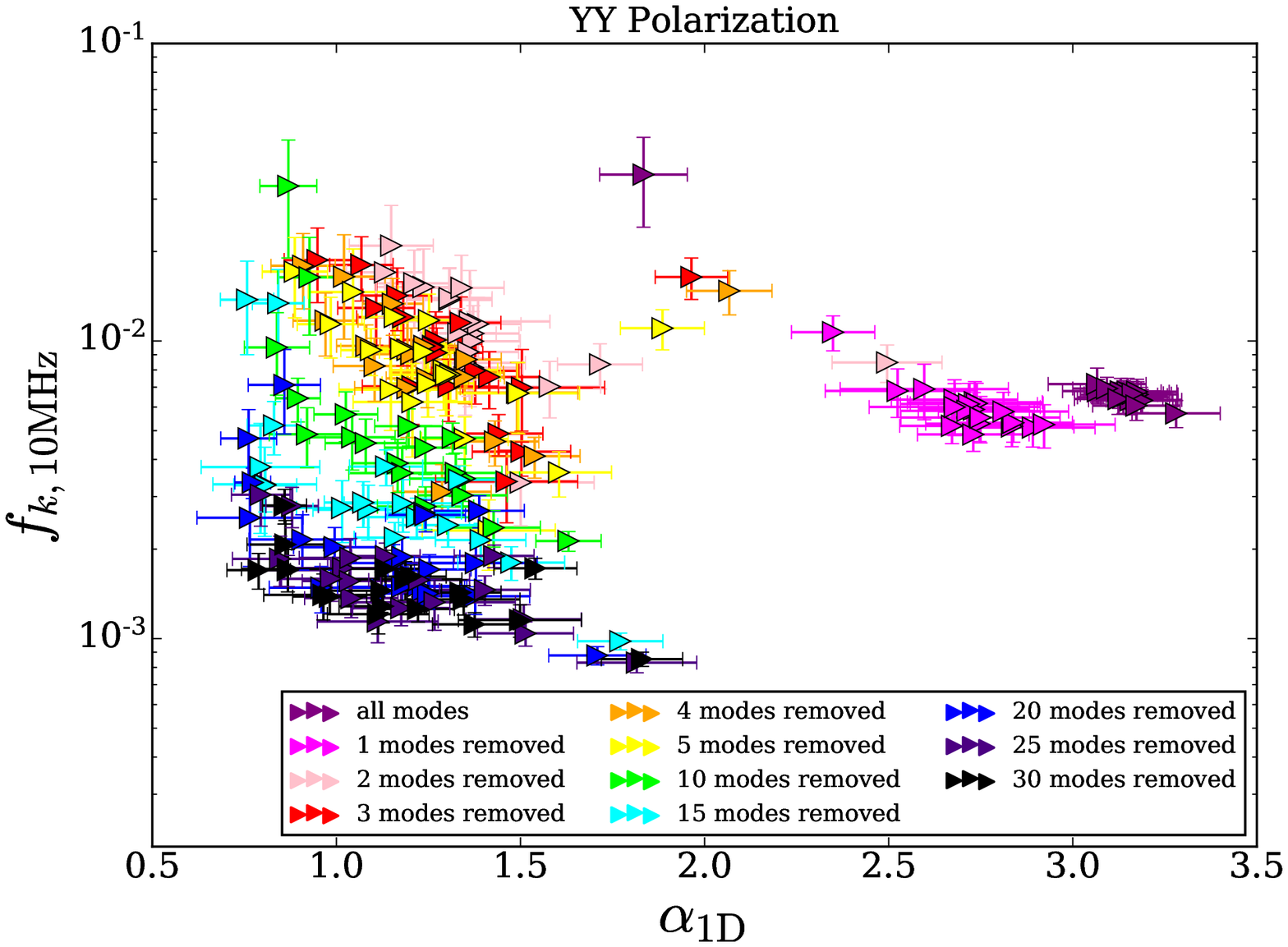}\par
\includegraphics[width=0.48\textwidth]{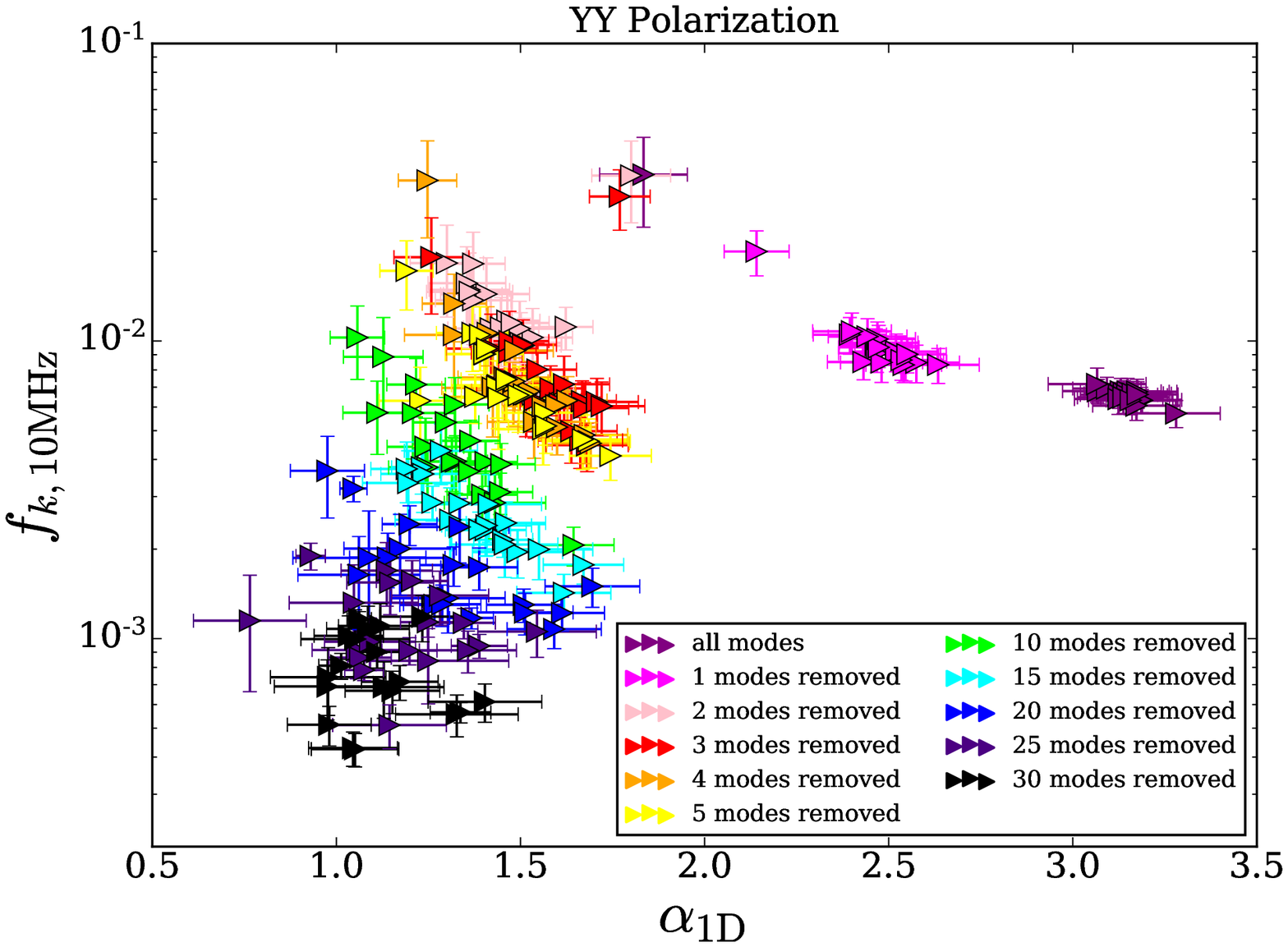}\par
\end{multicols}
\vspace*{-1.05cm}
\begin{multicols}{2}
\includegraphics[width=0.48\textwidth]{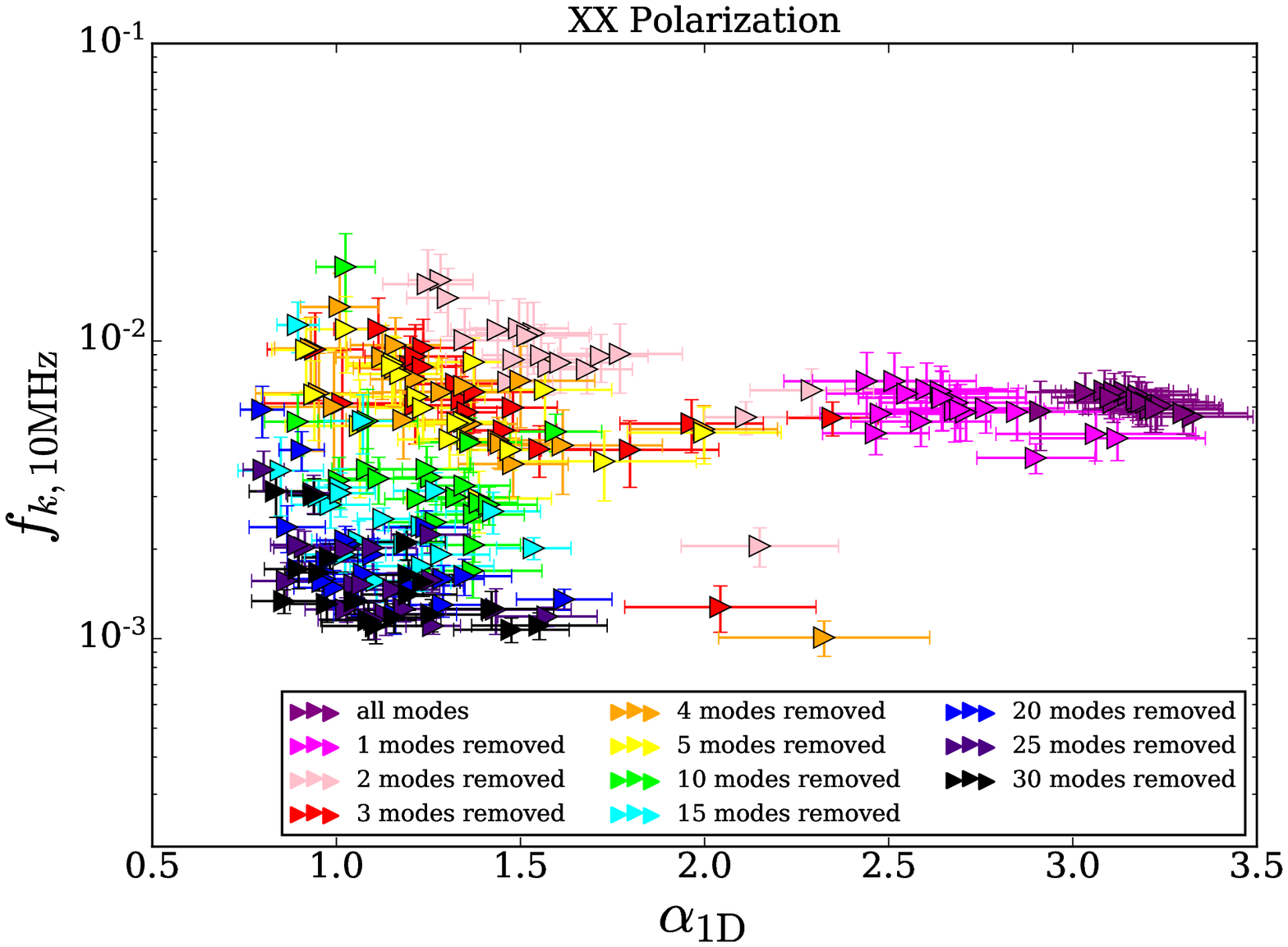}\par
\includegraphics[width=0.48\textwidth]{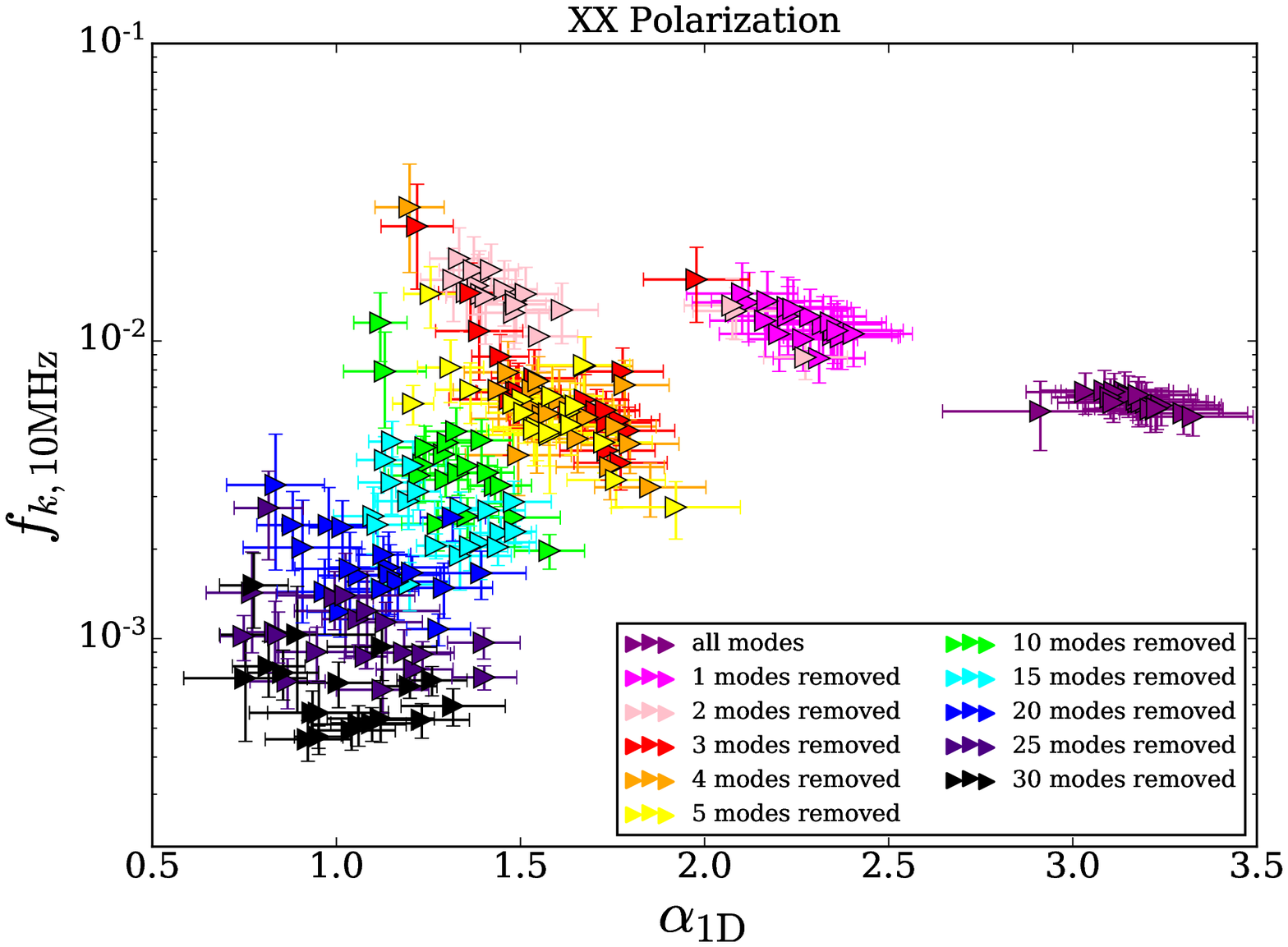}\par
\end{multicols}
\vspace*{-0.3cm}
\caption{The best-fit spectral index and the knee frequency of temporal power spectrum density with cross-correlation (left panels) and auto-correlation (right panels) SVD subtraction. The results with different levels of SVD subtraction are shown with different colors.}
\label{alpha_fk}
\end{figure*}

Besides the auto-correlation, we also apply the SVD subtraction to the cross-correlations of the different feeds. The 19 feeds are grouped into five east-west rows observing five different stripes of the sky. The same stripe of sky is observed in turn by multiple feeds sitting in the same row.
We use one feed in each row (Feed 1, Feed 4, Feed 7, Feed 12 and Feed 18, respectively) as $d^{\rm c}_{A}$ and the rest of the feeds in the same row as $d^{\rm c}_{B}$. Before estimating the cross-correlation SVD, the time stream data of two feeds are shifted in the time axis to make the pointing aligned. The singular values of the cross-correlation SVD are shown in the bottom panels of Figure~\ref{Smode}. To compare with the auto-correlation, we plot the square root of the original values, $\sqrt{{\rm diag}\{\Lambda_\times\}}$. The singular values of the cross-correlations decrease more rapidly to the noise floor with $\lesssim 10$ modes. Unlike the auto-correlations, cross-correlations are much less affected by correlated instrumental noise or gain variations, and dominated by sky signal. So this result means that the sky correlation can be subtracted down to the thermal noise fluctuations with $\lesssim 10$ modes. For the auto-correlations, more modes need to be subtracted to reduce the correlations to the thermal noise level. This indicates that in the auto-correlations there are probably more feed-dependent correlations remaining, which contribute to the 1/f noise of the system.

\section{Results}
\label{sec:Results}
\subsection{Temporal Power Spectrum}

In the temporal power spectrum analysis, we re-bin the frequency channels of the SVD-modes-subtracted data to 10 MHz resolution, corresponding to $\sim$ 50 Mpc at 1000 MHz (redshift $\sim$ 0.42). This reduces the white noise power level, but the 1/f noise is affected differently. As we noted earlier, the $f_{k}$ value depends on the channel width, and the re-binning pushes it to detectable range. The reduction of the 
frequency resolution would smooth out the fluctuations along the line of sight (LoS) on scales smaller than $\sim 20$ $h^{-1}$Mpc at $z \sim 0$, but these small-scale modes are not the focus of cosmological studies. After the re-binning we have 15 frequency channels across a bandwidth of 150 MHz. The temporal power spectrum is estimated for each feed and polarization. Finally, the mean temporal power spectra averaged across the $15$ frequency channels 
are presented. The power spectrum error is estimated from the variations across the $15$ frequency channels.

We estimate the temporal power spectrum from the time stream data, the results for feed 17, Feed 7, Feed 1, Feed 3 and Feed 11 are shown for both before and after cleaning in Figure~\ref{tps_cov} with the cross-correlation SVD modes, and in Figure~\ref{tps} with the  auto-correlation SVD modes. 
As can be seen from these figures, the results for the different feeds are remarkably similar with each other, 
and so are the other feeds which are not shown here. For all feeds, significant 1/f-type noise in the power spectrum is clearly visible. Even for the all-modes data, which is influenced mostly by the foreground and gain variations, its temporal power spectrum is still in 1/f shape. The power spectra at extreme high and low time scales are dominated by white noise, while the 1/f-type power dominates in the middle range.

The measured mean temporal power spectra with different number of SVD modes subtracted are shown with the error bars in the same figures in different colors.
Obviously, the overall power is reduced with the SVD subtraction, but there is still correlation power left at lower $f$.
The dashed lines show the best-fit temporal power spectrum model using Eq.~(\ref{psd}).

There is a bump at $f\sim 0.01$ Hz in the temporal power spectra of the XX-polarization data of Feed 1 with lower level of SVD subtraction (see lower-middle panel in Figure~\ref{tps_cov} and Figure~\ref{tps}). 
However, the bump is only found in the XX polarization of Feed 1 time stream data and remains in the data up-to $30$ cross-correlation SVD mode subtraction, but can be removed with $\sim 10$ auto-correlation SVD mode subtraction. The bumps are also identified in the analysis of the data from the other days. In order to investigate whether such bump is due to part of the bad data during the observation, we look into subsets of the data. We split the flagged time stream data into three subsets, each has one-third of the length of time of the original data. We run our pipeline with these subsets. For each subset calculation, the bumps are identified again at $f\sim 0.01$ Hz and the measured mean temporal power spectra with more than 15 modes removed are almost the same with those in Figure~\ref{tps_cov} and Figure~\ref{tps}. It seems that the bumps arise from the oscillation of the data across the whole time.

There is a plateau at $f \lesssim 10^{-4}$ Hz (corresponding to the time scales $\gtrsim 10^4 $ s ) for all feeds, with both
cross and auto-correlation SVD subtraction. 
Such truncation of the power spectrum at lower $f$-end might indicate a maximum correlation length of 1/f-type correlation. However, this measurement of the lower $f$ end may also be affected by the finite length of available data, which lasts only about $5\times 10^4 s$.
With the number of subtracted SVD modes increases from 0 to 30, the plateau expands from $\lesssim$ 4 $\times$ 10$^{-5}$ Hz to $\lesssim$ 10$^{-4}$ Hz.

\begin{figure}
    \centering
    \includegraphics[width=0.48\textwidth]{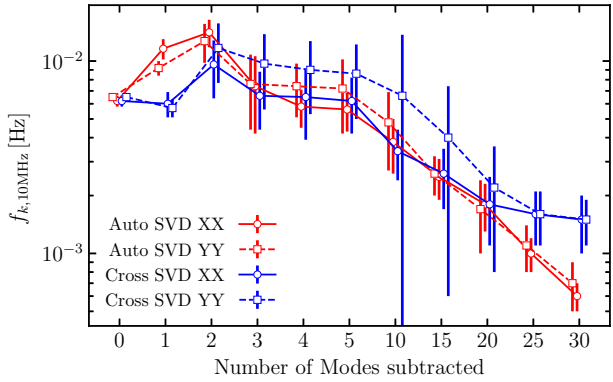}
    \caption{
    The best-fit knee frequency of the temporal power spectrum density
    with different levels of SVD subtraction.
    The results are averaged over all of the feeds and the r.m.s across the feeds are plotted as error bars, but with bad feeds excluded. The results with auto-correlation SVD subtraction are shown in red
    and the the cross-correlation SVD subtraction results are in blue. 
    The XX polarization is shown using circles with solid lines and
    the YY polarization is shown using squares with dashed lines.
    }\label{fkac}
\end{figure}

We then fit the power spectrum using Eq.(\ref{psd}). Figure~\ref{alpha_fk} shows the best-fit spectral index and the knee frequency of the time stream data averaged over the feeds, with the cross (left panels) and auto-correlation (right panels) SVD subtraction, respectively. The results with different levels of SVD subtraction are shown with different colors.
With the number of SVD modes subtraction increases from $0$ to $30$, 
the spectral index drops from $\sim 3.0$ to $\sim 1.0$ for both the cross and auto-correlation SVD subtraction. 
The knee frequency behaves differently for the auto- and cross- correlations.
For cross-correlation SVD subtraction, the knee frequency is reduced 
from $\sim10^{-2}$ to $\sim2\times10^{-3}$ with $20$ modes subtracted, and stays at the same level as more modes are subtracted. 
However, for auto-correlation, 
the knee frequency goes down to $\sim$ 6 $\times$ 10$^{-4}$ Hz with another $10$ more modes subtracted. 

The differences can be seen more directly in Figure~\ref{fkac}, where knee frequency averaged over all 19 feeds are shown 
in red for auto-correlation SVD subtraction, and in blue for cross-correlation SVD subtraction.
As the cross-correlation SVD modes subtraction removes the correlations from the sky variations, the remaining correlations after the subtraction of 20 modes are dominated by the system-induced 1/f noise, and the knee frequency is $\sim2\times10^{-3}$ Hz. Such system-induced 1/f noise can be further reduced with  auto-correlation SVD modes. With another $10$ modes subtracted, the knee frequency is reduced to $\sim 6 \times 10^{-4}$ Hz, indicating that the system-induced 1/f-type variations are well under the thermal noise fluctuations over 
$1600$ seconds time scales. 
In the rest of the analysis, we focus on the results with auto-correlation SVD subtraction.

\begin{figure}
    \centering
    \includegraphics[width=0.49\textwidth]{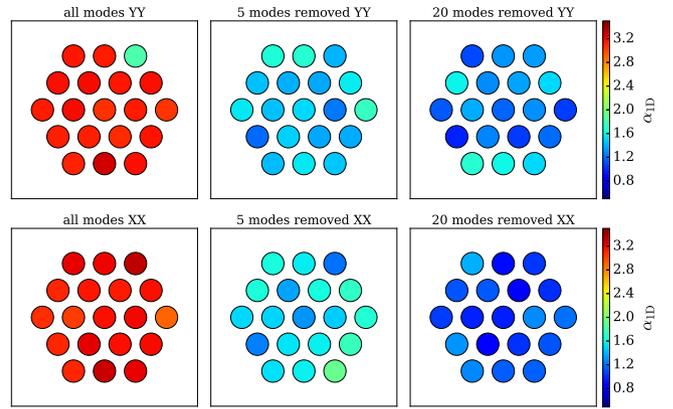}
    \caption{The distribution on feeds of the spectral index of the 1/f-type noise, from the fitting of the temporal power spectra. The XX and YY polarizations are shown in the lower and upper subpanels. The deployment and label of feeds are the same as those in Figure~\ref{feeds}.}
    \label{alpha_beams}
\end{figure}

\begin{figure}
    \centering
    \includegraphics[width=0.49\textwidth]{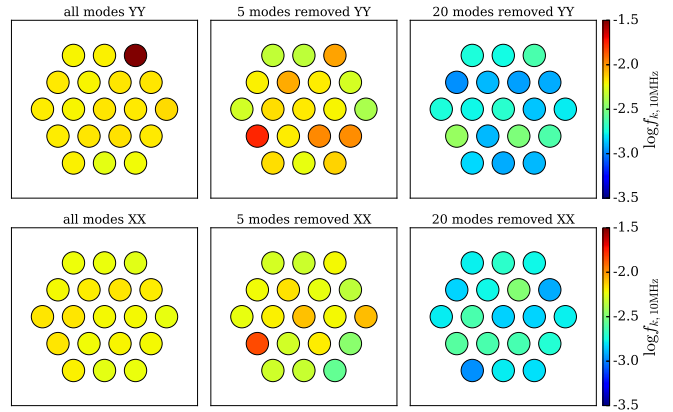}
    \caption{Same as Figure~\ref{alpha_beams}, but for the knee frequency.}
    \label{fk_beams}
\end{figure}

\begin{figure*}
    \centering
    \includegraphics[width=0.96\textwidth]{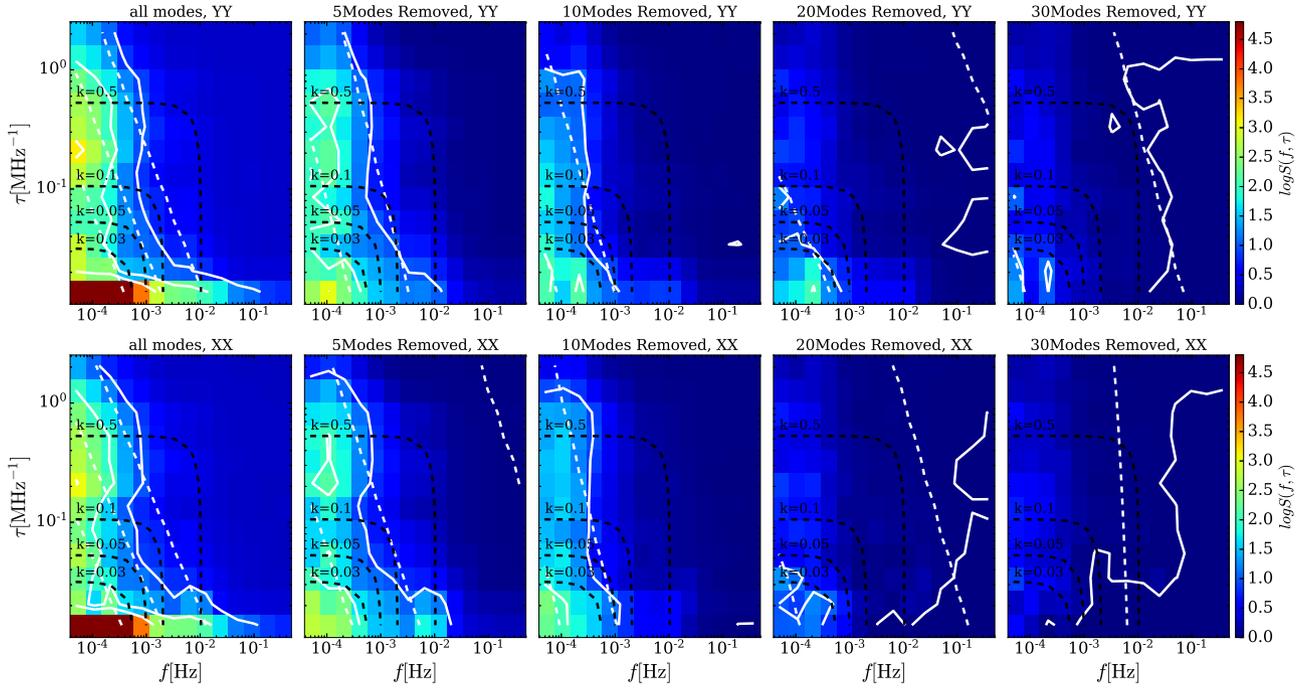}
    \caption{The 2D power spectra for Feed 1. Upper panels: YY polarization. Lower panels: XX polarization. The levels of 10$^{0}$, 10$^{1}$ 10$^{2}$ and 10$^{3}$ are shown by solid-white lines. The dashed-white lines show the fitted 2D power spectra's same levels as measurements. }
    \label{2dps}
\end{figure*}

\begin{table*}
	\centering
	\caption{The mean value of 10MHz temporal power spectra and 2D power spectra fitting parameters across all feeds except Feed 16 YY and Feed 9 XX, from auto-correlation foreground subtraction. The errors are the r.m.s. of the fitting values across all feeds except Feed 16 YY and Feed 9 XX.}
	\label{alpha_fk_beams_table}
	\begin{tabular}{|c|cc|cc|cc|cc|}
		\hline
		Data & $ f_{k,10\rm MHz}\times 10^2$ & & $\alpha_{\rm 1D}$ & & $\alpha_{\rm 2D}$ & & $\beta_{\rm 2D}$\\
		\hline
		 & XX & YY & XX & YY & XX & YY & XX & YY\\
		\hline
		all modes data       & 0.62 $\pm$ 0.04 & 0.65 $\pm$ 0.03 & 3.16 $\pm$ 0.09 & 3.15 $\pm$ 0.05 & 1.45 $\pm$ 0.06 & 1.50 $\pm$ 0.20 & 0.40 $\pm$ 0.03 & 0.43 $\pm$ 0.02\\
         20 modes removed & 0.18 $\pm$ 0.05 & 0.17 $\pm$ 0.07 & 1.08 $\pm$ 0.14 & 1.30 $\pm$ 0.20 & 0.77 $\pm$ 0.11 & 0.78 $\pm$ 0.12 & 0.65 $\pm$ 0.07 & 0.58 $\pm$ 0.04\\
        25 modes removed & 0.10 $\pm$ 0.02 & 0.11 $\pm$ 0.03 & 1.05 $\pm$ 0.18 & 1.17 $\pm$ 0.17 & 0.72 $\pm$ 0.10 & 0.74 $\pm$ 0.10 & 0.74 $\pm$ 0.08 & 0.63 $\pm$ 0.08\\
        30 modes removed & 0.06 $\pm$ 0.01 & 0.07 $\pm$ 0.02 & 1.03 $\pm$ 0.15 & 1.12 $\pm$ 0.12 & 0.63 $\pm$ 0.10 & 0.64 $\pm$ 0.11 & 0.84 $\pm$ 0.08 & 0.72 $\pm$ 0.10\\
        
		\hline
	\end{tabular}
\end{table*}

\begin{table*}
	\centering
	\caption{The mean value of 10MHz temporal power spectra and 2D power spectra fitting parameters across all feeds except Feed 16 YY and Feed 9 XX, from cross-correlation foreground subtraction. The errors are the r.m.s. of the fitting values across all feeds except Feed 16 YY and Feed 9 XX.}
	\label{alpha_fk_beams_table_cov}
	\begin{tabular}{|c|cc|cc|cc|cc|}
		\hline
		Data & $ f_{k,10\rm MHz}\times 10^2$ & & $\alpha_{\rm 1D}$ & & $\alpha_{\rm 2D}$ & & $\beta_{\rm 2D}$\\
		\hline
		 & XX & YY & XX & YY & XX & YY & XX & YY\\
		\hline
		all modes data       & 0.62 $\pm$ 0.04 & 0.65 $\pm$ 0.03 & 3.16 $\pm$ 0.09 & 3.15 $\pm$ 0.05 & 1.45 $\pm$ 0.07 & 1.50 $\pm$ 0.20 & 0.40 $\pm$ 0.03 & 0.43 $\pm$ 0.02\\
         20 modes removed & 0.18 $\pm$ 0.07 & 0.22 $\pm$ 0.14 & 1.11 $\pm$ 0.18 & 1.12 $\pm$ 0.23 & 0.66 $\pm$ 0.09 & 0.67 $\pm$ 0.08 & 0.63 $\pm$ 0.05 & 0.59 $\pm$ 0.06\\
        25 modes removed & 0.16 $\pm$ 0.05 & 0.16 $\pm$ 0.05 & 1.12 $\pm$ 0.18 & 1.15 $\pm$ 0.26 & 0.64 $\pm$ 0.09 & 0.66 $\pm$ 0.08 & 0.65 $\pm$ 0.06 & 0.61 $\pm$ 0.05\\
        30 modes removed & 0.15 $\pm$ 0.05 & 0.15 $\pm$ 0.04 & 1.14 $\pm$ 0.19 & 1.15 $\pm$ 0.26 & 0.64 $\pm$ 0.09 & 0.65 $\pm$ 0.08 & 0.66 $\pm$ 0.06 & 0.62 $\pm$ 0.05\\
		\hline
	\end{tabular}
\end{table*}

In Figure~\ref{alpha_fk} we can also find that the best-fit values of most of the feeds are similar except the results from Feed 9 XX and Feed 16 YY (the outliers in Figure~\ref{alpha_fk}). Furthermore we plot the FAST L-band Array of 19 feed-horns again in Figure~\ref{alpha_beams} and Figure~\ref{fk_beams}, colored with the value of the spectral index and
the knee frequency, respectively. 
Except Feed 16 and Feed 9, the 1/f-type noise behavior for most of the feeds are similar.
It seems that there are large fluctuations in the gain of Feed 16 YY and Feed 9 XX during the observation. Such gain fluctuations can be reduced by removing a few more SVD modes, however, it does indicate that corresponding receiver channels have some abnormal behaviors. We will exclude the Feed 9 XX and Feed 16 YY polarization in the following statistical analysis. In Table ~\ref{alpha_fk_beams_table} and Table ~\ref{alpha_fk_beams_table_cov} we list the mean and r.m.s. values of the fitting parameters for the 10MHz-resolution band of all feeds except Feed 16 and Feed 9, for the auto- and cross-correlation SVD subtraction respectively. 

To check how the results depend on the size and the resolution of the data cube, we have splitted the time stream data into 3 subsets, each has equal time length. The analysis for each subset give results similar to the whole set. We also re-bin the data into time resolution of 0.2s and 8s and compare the analysis results. The differences appear mostly in the first 5 SVD modes.  After the first 5 SVD modes are removed the results are similar, though with the lower time resolution the error is larger. These tests show the results of our analysis are quite robust.

\subsection{2D Power Spectrum}

The 2D power spectrum density is estimated by Fourier transforming the data along both the time and frequency axes. The Fourier conjugate variable $\tau$ for the frequency $\nu$ is called {\it delay}.  
The data we use here has time and frequency resolution of 1.0 s and 0.2 $\MHz$. The results
for Feed 1 are shown in the contour plots in Figure~\ref{2dps} as an example. From left to the right panels, it shows the 2D power spectra with all-modes data and data with 5, 10, 20 and 30
auto-correlation SVD modes subtracted. The results for the two polarizations are shown in the upper and lower subpanels, respectively. 
The contour levels of 10$^{0}$, 10$^{1}$ 10$^{2}$ and 10$^{3}$ are shown by solid-white lines. 
The dashed-white lines show the fitted 2D power spectra of Eq.~(\ref{psd2d}) at the same levels as the measurements. 

The power spectra are peaked at the low-$\tau$ end, which indicates a strong correlation across the whole frequency channels. There are also weak correlations at smaller delay intervals, especially at $\tau\sim 0.2\, {\rm MHz}^{-1}$, which may be due to the frequently seen chronical fixed-frequency RFI. To investigate the impact of such RFI, we ignore the mask in frequencies and estimate the correlation power of such RFI in frequency with our 2-D power spectrum estimator. We found a peak at the same correlation scales ($\tau\sim 0.2\, {\rm MHz}^{-1}$) with much stronger power. Although the correlation power at $\tau\sim 0.2\, {\rm MHz}^{-1}$ have been significantly reduced after masking such RFI and filling the gaps by interpolation, the correlations persists in the data. The correlations left may arise from the residual of the RFI and un-identified weak RFI, which may have similar correlations to the ones removed. 
However, these strongly correlated components can be removed by subtracting 
the first several SVD modes. As shown in the second row of Figure~\ref{2dps}, 
after $5$ SVD modes subtracted, the power at low $\tau$-end is already highly reduced, and the correlations across the whole frequency channels are suppressed. With 20 SVD modes subtracted, the strongest contamination is constrained in low $f$-$\tau$ space. However, subtraction of fewer SVD modes will still leave frequency correlations in the estimated power spectra. As discussed with the temporal power spectrum analysis, the first $20$ modes are dominated by the external correlations from sky. 

We also show the corresponding cosmological scales projected to the $f$-$\tau$ space
as the dashed-black lines in Figure~\ref{2dps}. 
The parallel and vertical cosmological scale k$_{\parallel}$ and k$_{\perp}$ 
can be obtained by:
\begin{eqnarray}
    k_{\parallel} = 2\pi \frac{H(z)}{c}\frac{\nu^2_{\rm obs}}{\nu_{0}}\tau,  \ \ \ 
    k_{\perp} = 2\pi\frac{f}{\chi(z)u},
    \label{cosmo_scale}
\end{eqnarray}
where $\nu_{0}$ = 1420 MHz is the rest frame frequency for \hi emission; 
$c$ is the speed of light; $\nu_{\rm obs}$ is the observing frequency and $u$ is the speed of scanning; $\chi(z)$ is the comoving distance at redshift $z$. The cosmological parameters used here are from \citet{2018arXiv180706209P}.  
We adopt $\nu_{\rm obs} = 1.1$ GHz and $u = 0.25\, {\rm arcmin/s}$ for FAST drift scan experiment. With 20 SVD modes subtracted, the strongest contamination is constrained inside k $=$ 0.05 h/\Mpc.

\begin{figure}
    \centering
    \includegraphics[width=0.46\textwidth]{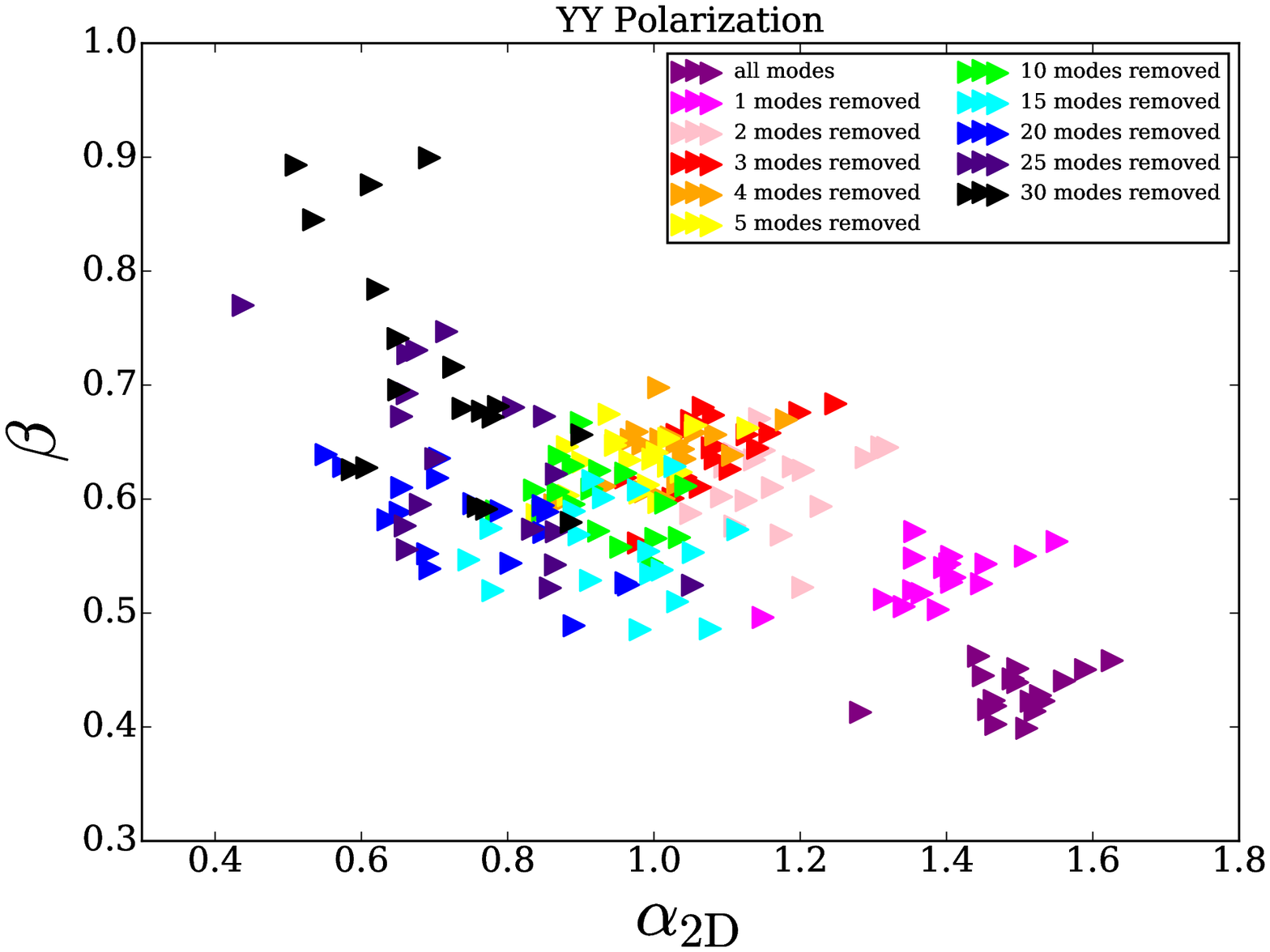}
    \includegraphics[width=0.46\textwidth]{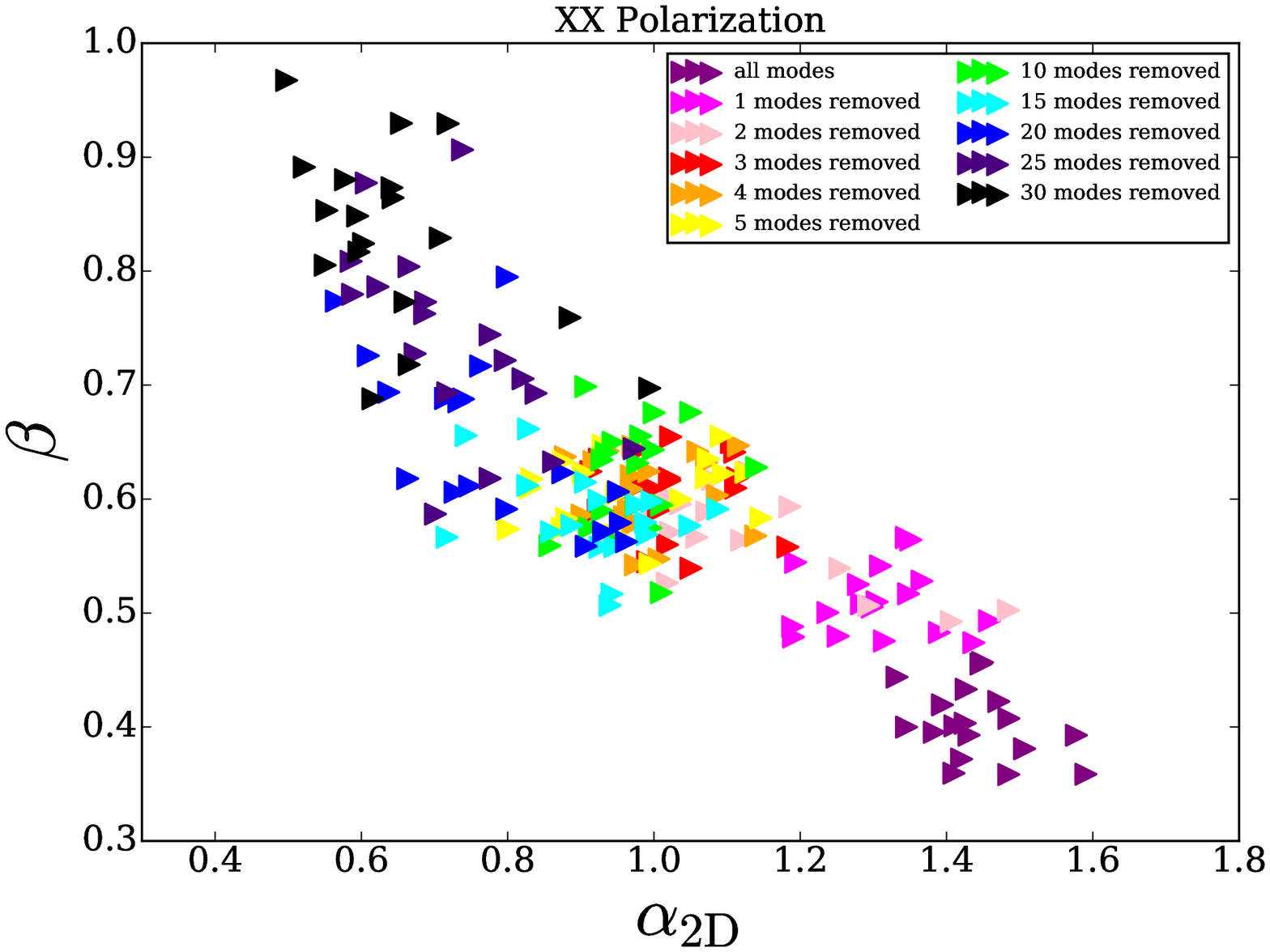}
    \caption{The spectral index $\alpha$ and $\beta$ of the 1/f-type noise, from the fitting to the 2D power spectra. The XX and YY polarizations are shown in the lower and upper panels.}
    \label{alpha_beta}
\end{figure}

By fitting Eq. ~(\ref{psd2d}), we obtain the 
best-fit values of $\alpha$ and $\beta$, and study how they change with different levels of SVD mode subtraction. 
The mean value of the fitting parameters, as well as the r.m.s. over all feeds are 
listed in the right two columns in Table ~\ref{alpha_fk_beams_table}, again we exclude Feed 9 XX and Feed 16 YY when calculate the statistical values. The results are also shown in Figure~\ref{alpha_beta}, with the upper and lower sub-panel for
the YY and XX polarizations respectively. Each marker shows the best-fit value for one feed, and different SVD modes 
subtractions are shown with different colors as indicated in the legend. 

The best-fit $\beta$ values for the data without SVD subtraction are around $\sim 0.4$,
indicating a strong correlation across frequencies. 
The $\beta$ value increases with more modes subtracted, reaching
$\beta \sim 0.8$ for $30$ SVD modes subtracted, indicating a much weaker
correlation across frequency channels. 
For the spectral index parameter $\alpha$, it goes down from $\sim1.3$ to 
$\sim 0.6$ with the SVD subtraction from $0$ to $30$ modes, indicating weaker 1/f-type noise fluctuations.

While we have mainly presented the results from the analysis of the data from a single day's observation, it is worth mentioning that we also performed analysis for data from other days. Although on each day the observed sky is slightly different, and there are some changes in observation parameters (noise-diode power, time and frequency resolutions), all results are quite consistent and similar, and we obtain similar values for fitting parameters ($\alpha$, $\beta$ and $f_{k}$) after the first 20 SVD modes are removed.

\subsection{Gain Calibration}

\begin{figure}
    \centering
    \includegraphics[width=0.46\textwidth]{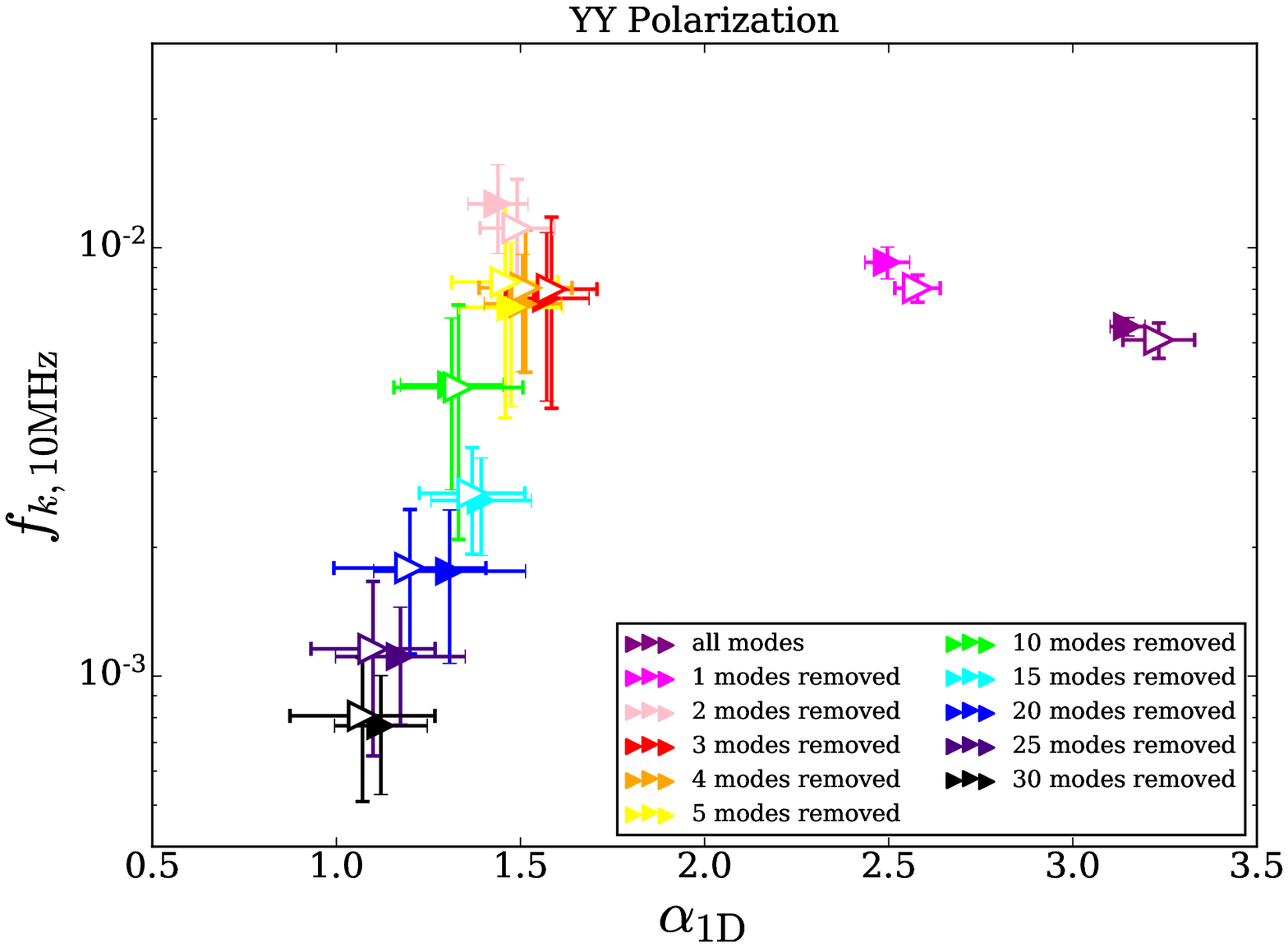}
    \includegraphics[width=0.46\textwidth]{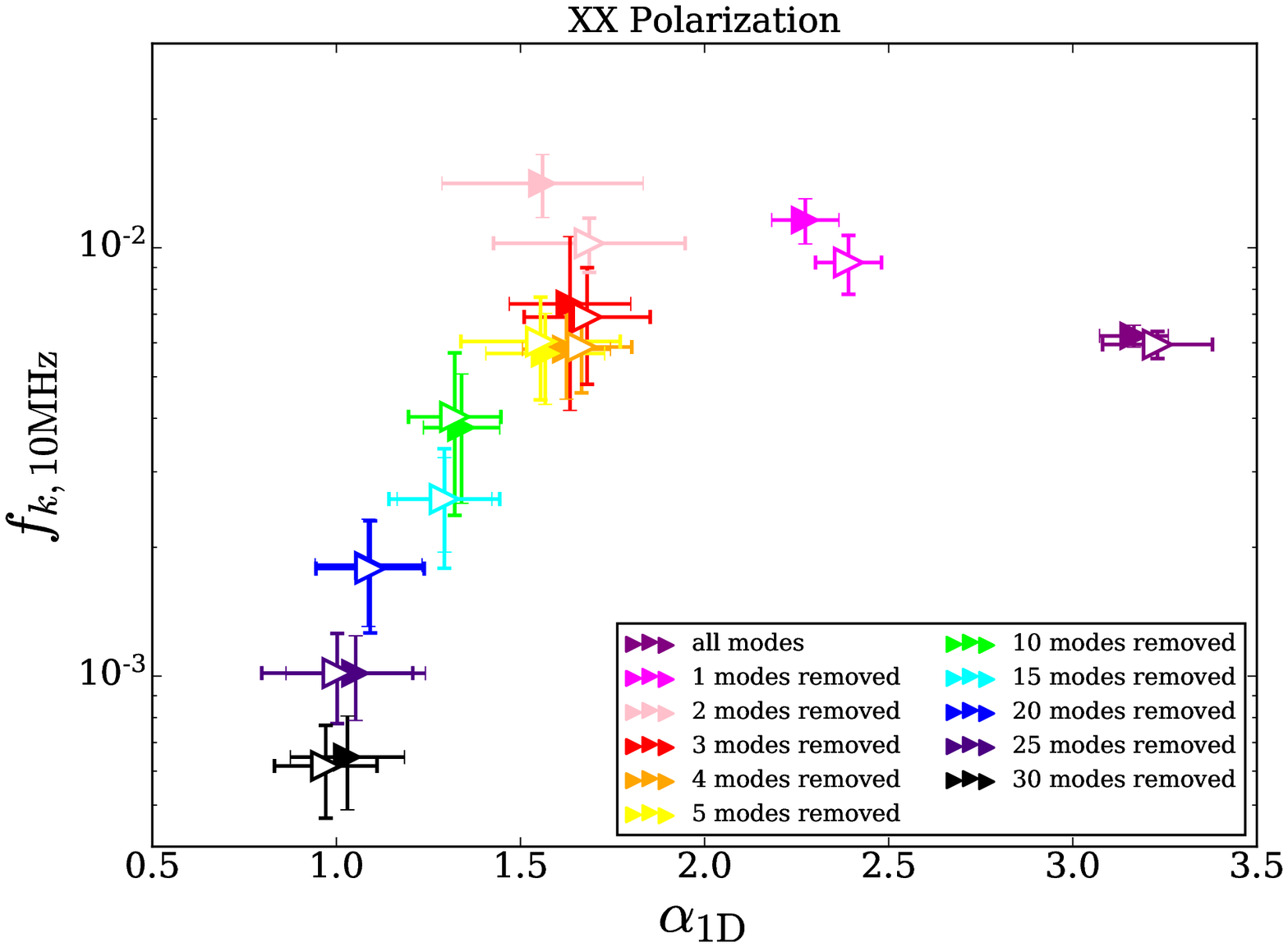}
    \caption{Comparison of the knee frequency and spectral index from temporal power spectrum density for the gain-uncalibrated data and the gain-calibrated data. The filled/empty points represent the data without/with gain calibration.}
    \label{alpha_fk_gaincali_compare}
\end{figure}

In the analysis above, we estimated the 1/f noise using the real-time uncalibrated data. We have also tried to apply the analysis on the real-time calibrated data, for if the 1/f noise comes mainly from the fluctuations of the gain, it may be mitigated after the gain calibration. The built-in noise diode of the FAST receiver is used as a real-time calibrator to estimate the gain variations, $G(t, \nu)$. The noise diode is turned on for 1 second in every 8 seconds, with an amplitude of 1 K, which is a small fraction of the system temperature. The noise diode modulation is used to calibrate the time variations of the overall gain $\bar{G}(t)$ (i.e. average over the whole frequency band), frequency-dependent variations are not corrected in this procedure. We then carry out the same analysis using the calibrated data. However, we find that the result is little affected. Figure~\ref{alpha_fk_gaincali_compare} compares the best-fit spectral index and the knee frequency of the real-time uncalibrated data (shown as filled points in the figure) and calibrated data (shown as empty points) obtained with the auto-correlation SVD subtraction, and we see the results are very similar. As the knee frequency is determined by both the 1/f and white noise, the results indicate that at least for the simple calibration procedure outlined above, the 1/f noise is not reduced. Either our calibration is not sufficiently accurate, or the 1/f noise is not primarily associated with the frequency-independent gain variations. The precision of the calibration procedure is limited, due to noise or fluctuations in the calibrator signal. It is also possible that the 1/f noise does not arise simply as the time variations of the frequency-independent part of the receiver gain which our calibration procedure corrects, but may have more complicated frequency dependence.

\subsection{Impact on Power Spectrum Measurement}
We use a simple simulation to assess how such 1/f noise can affect cosmological measurement.
We start with a set of lognormal mock maps as the input \hi signal, the time stream data are generated by scanning the mock maps. We assume $100$-hours observation time for each feed and split it into $25$ drift scans, 
each with a shift of $10.835\,{\rm arcmin}$ in Dec. The simulated time stream data   
cover $60^\circ$ in R.A. and $\sim 4.6^\circ$ in Dec. 
We generate the simulated noise for the following cases: (CASE 0) $20{\rm K}$ white noise;
(CASE 1) $20\,{\rm K}$ white noise and 1/f-type fluctuations, using the noise parameters after subtracting 20 SVD modes from 
Table ~\ref{alpha_fk_beams_table}; (CASE 2) Same as above, except after subtracting 30 modes.

\begin{figure}
    \centering
    \includegraphics[width=0.46\textwidth]{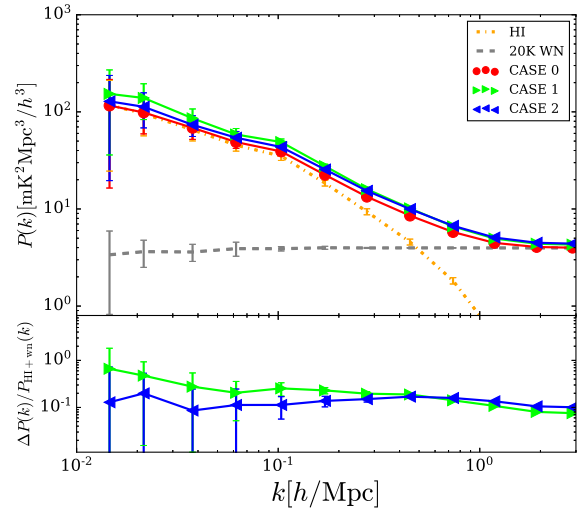}
    \caption{Upper panel: The power spectra from simulated sky with different noise injected. The orange dot-dashed line and grey dashed line are the power spectra of \hi sky and white noise, respectively. Lower panel: the relative difference between the power spectra of \hi sky with only white noise injected and 1/f-type noise injected.}
    \label{ps_1d_simulation}
\end{figure}

We calculate the power spectra of the simulated sky with different noise models, and present the results in Figure ~\ref{ps_1d_simulation}. The upper panel shows the power spectra of each case, the orange dot-dashed line and grey dashed line are the \hi and
white noise power spectrum, respectively. 
The lower panel presents the relative difference.
It shows that the non-physical structures induced by 1/f-type noise enlarge the power spectrum. 
With stronger 1/f noise, the values of the power spectra are larger. 
For the 1/f noise with $f_{k} \sim$ 1.8 $\times$ 10$^{-3} \Hz$, $\alpha \sim$ 0.8 and 
$\beta \sim$ 0.6 (parameters from 20 SVD modes removed), the relative difference is $\sim$ 0.25. 
For the 1/f noise with $f_{k} \sim$ 6 $\times$ 10$^{-4} \Hz$, $\alpha \sim$ 0.65 and $\beta \sim$ 0.8 (parameters from 30 SVD modes removed), the relative difference is $\sim$ 0.1. 
In other words, the 1/f-type noise with those parameters will enlarge the power spectrum by 10 percent. For FAST drift-scan \hi survey, the 1/f noise can be suppressed by the SVD method 
to a low level, but the influence of 1/f noise can not be ignored. 

\section{Discussions}
\label{sec:Discussions}

Besides the 1/f noise in the receiver system, the natural variations of the sky signal in the drift scan also contribute a power-law component to the fluctuations in the time stream data \citep{2015MNRAS.454.3240B}. This is however a natural fluctuation in the sky signal itself. In Sec.~\ref{sec:Results}, we compared the analysis for 1/f noise from time stream data cleaned with the cross-correlation and the auto-correlation SVD modes subtraction. Note that for the cross-correlations, the receiver noise is largely cancelled, yet we still see nearly the same power spectrum before subtracting the first 20 SVD modes, this indicates 
that the 1/f component from the first 20 SVD modes are dominated by the external correlations from the sky. 

In order to evaluate the sky component, we carry out numerical simulations. We generate a mock foreground with the global sky model (GSM) \citep{2008MNRAS.388..247D,2017MNRAS.464.3486Z,Huang:2018ral} and convert it into time stream data by scanning the mock maps. The data covers a frequency range from $1050\,{\rm MHz}$ to $1150\,{\rm MHz}$ with a resolution of $1\,{\rm s}$ in time and $1\,{\rm MHz}$ in frequency. We have produced two sets of simulations, one is foreground only, and the other is foreground + noise. The noise here consists of $20{\rm K}$ white noise and 1/f-type fluctuations simulated using the measured noise parameters after 20 modes subtracted, as given in Table ~\ref{alpha_fk_beams_table}.

\begin{figure}
    \centering
    \includegraphics[width=0.46\textwidth]{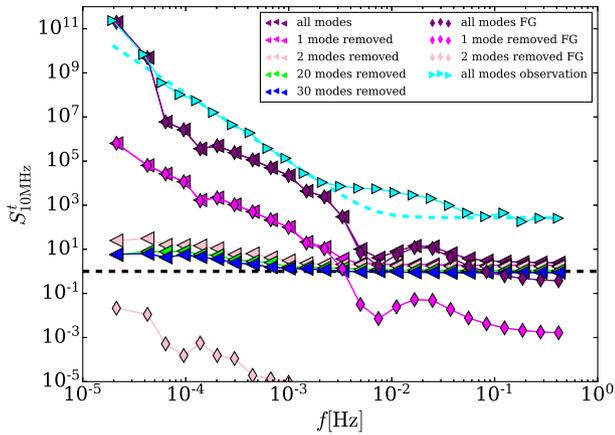}
    \caption{The frequency averaged temporal power spectra of simulated time stream data cleaned with auto-correlated SVD modes. The left-pointing triangle refers to the foreground + noise case and the diamond shows the foreground case. We also present the temporal power spectrum of the all-modes data from observation in right-pointing triangle. The thermal noise part of these spectra are normalized to $1$. The dashed-black line shows the value of unit.}
    \label{foreground_simulation}
\end{figure}

We run our data processing pipeline to measure the temporal power spectra for the above two cases, and the results are shown in Fig. ~\ref{foreground_simulation}.  The left-pointing triangle and the diamond symbols show the foreground+noise case and the purely foreground case respectively. For comparison, the temporal power spectra of the all-modes data from observation is shown as right-pointing triangles. As we can see from Fig. ~\ref{foreground_simulation}, 
the ``all modes" power spectrum of the simulation matches the observation
at $f<3\times10^{-3}$ Hz, which shows that
the foreground signal makes the major contribution to the 1/f-type correlations in the data before SVD cleaning. The simulated power spectrum is suppressed at shorter time scales than for the actual observation, which is expected given the limited angular 
resolution of the GSM mocks. 

However, in the simulation, the foreground power can be reduced more significantly with SVD mode subtraction. For the simulation, after only 2 modes are subtracted, the foreground power spectrum 
is well below the the noise level, and the system induced 1/f noise starts to dominate.  Subtractions of additional modes do not reduce the 1/f noise more significantly.

For the real observation data, $\sim 20$ modes are needed to reduce the foreground power spectrum down to the noise level, which are many more than the number of modes needed in the simulation. This is not surprising, as the simulations necessarily made simplifications to the instrument effect. Inclusion of more instrumental effects, e.g. the antenna primary beam, polarization leakage, etc., or more sophisticate treatment can significantly increase the degree of freedom of foreground
\citep{2020arXiv201002907C,2020arXiv201110815M}. 
Thus, in the real data, there are always many more large modes, which arise not purely from receiver noise, but also from the coupling of the sky signal with the system response.

Comparing our results with a similar analysis for a MeerKAT observation, many more SVD modes need to be subtracted in order to remove the correlations induced by the sky \citep{2020arXiv200701767L}. A possible reason of this is that in that case, the observation was carried out with a single pointing at the South Celestial Pole (SCP), so the naturally existing correlations are much simpler, its correlations can be well removed with 2 SVD modes subtracted. In this work, we observed a long strip of sky, which may harbour more correlations and require more SVD modes to be subtracted to remove the correlations arising from this. 

After the first $20$ modes have been subtracted, the remaining correlations in the data are dominated by the system induced 1/f noise, and the knee frequency averaged across feeds is reduced to $\sim 1.8 \times 10^{-3}$ Hz with r.m.s. of $0.5\times 10^{-3}$ Hz. The system induced 1/f noise can be further reduced by subtracting
more modes estimated with the auto-correlation SVD. With $30$ modes subtracted, the knee frequency is reduced to $(6\pm1)\times10^{-4}$ Hz. These knee frequency values are of the same order of magnitude as those for the MeerKAT, 
which has $f_k = 3\times10^{-3}$ Hz at $20$ MHz frequency resolution \citep{2020arXiv200701767L}.
We do not expect the knee frequency to be exactly the same for different receiver systems, but it seems that the 1/f noise levels in these different systems are of comparable magnitude.

\section{Summary}
\label{sec:Summary}

In this work, we investigate the 1/f noise properties of the FAST receiver system using drift-scan data from an Intensity Mapping experiment. Both the temporal power spectrum density and the 2D power spectrum density are measured. All of the 19 feeds of the L-band system are analyzed for both polarizations. 
We select the data from a RFI-quiet part of the frequency band, $1.3\sim1.45$ GHz, to be processed by our pipeline. The data is re-binned to 1.0 s and 0.2 MHz resolution in time and
frequency. We process the raw data successively with flagging, bandpass normalization, singular value decomposition (SVD) and gap filling, before the final power spectrum estimation.

We obtain SVD modes from applying both auto-correlations and cross-correlations to the data set.
The results show that, by removing the strongest components, the 1/f-type correlations in the data
can be reduced significantly. 
The cross-correlation SVD subtraction analysis shows that the sky fluctuations
dominate the first $\sim20$ SVD modes. 
With first 20 modes subtracted, the knee frequency of the 1/f noise temporal power spectrum is reduced to 1.8 $\times$ 10$^{-4}$ Hz (for 10 MHz band), indicating that the system induced 1/f-type variations are well below the thermal noise fluctuations over 
5 hundred seconds time scales. The 2D power spectrum shows that the 1/f-type variations are restricted to a small region in the $f$-$\tau$ space and the correlations in frequency can be suppressed with SVD modes subtraction.

It is possible that the 1/f-type fluctuations can be further suppressed by calibrating against the real-time calibrator, though in our very simple treatment this is not obvious. In addition, a couple of destriping methods
have been proposed and tested 
\citep{2002A&A...387..356M,2004A&A...428..287K,2009A&A...506.1511K}, which may be applied to the FAST data. The destriping methods require the knowledge of the noise correlations. The noise power spectrum measurements presented in this work, 
as well as the best-fit noise model, provide good approximation of the 
noise correlations feature for the further destriping analysis of
FAST \hi intensity mapping. These techniques may be tried in the future to improve the result. 

\section*{Acknowledgements}
This work is supported by the National Key R\&D Program 2017YFA0402603, the Ministry of Science and Technology (MoST) inter-government cooperation program China-South Africa Cooperation Flagship project 2018YFE0120800, the MoST grant 2016YFE0100300, the National Natural Science Foundation of China (NSFC) key project grant 11633004, the Chinese Academy of Sciences (CAS) Frontier Science Key Project QYZDJ-SSW-SLH017 and the CAS Interdisciplinary Innovation Team grant (JCTD-2019-05), the NSFC-ISF joint research program No. 11761141012,  the CAS Strategic Priority Research Program XDA15020200, and the NSFC grant 11773034. Yichao Li acknowledge support from the South Africa National Research Foundation (NRF) through the South African Research Chairs Initiative (SARChI) Grant No. 84156 and the South African Radio Astronomy Observatory (SARAO)
group grant. Wenkai Hu and Guilaine Lagache are supported from the European Research Council (ERC) under the European Union's Horizon 2020 research and innovation programme (project CONCERTO, grant agreement No 788212) and from the Excellence Initiative of Aix-Marseille University-A*Midex, a French "Investissements d'Avenir" programme. Bo Zhang is supported by NSFC with grant no.11903056. Y.Z.Ma is supported by NRF with grant no.105925, 109577, 120378, 120385.

\section*{Data Availability}
The radio data analyzed in this work can be accessed by sending request to the FAST Data Centre or to the corresponding authors of this paper.

\bibliographystyle{mnras}
\bibliography{noise}

\begin{thebibliography}{}
\makeatletter
\relax
\def\mn@urlcharsother{\let\do\@makeother \do\$\do\&\do\#\do\^\do\_\do\%\do\~}
\def\mn@doi{\begingroup\mn@urlcharsother \@ifnextchar [ {\mn@doi@}
  {\mn@doi@[]}}
\def\mn@doi@[#1]#2{\def\@tempa{#1}\ifx\@tempa\@empty \href
  {http://dx.doi.org/#2} {doi:#2}\else \href {http://dx.doi.org/#2} {#1}\fi
  \endgroup}
\def\mn@eprint#1#2{\mn@eprint@#1:#2::\@nil}
\def\mn@eprint@arXiv#1{\href {http://arxiv.org/abs/#1} {{\tt arXiv:#1}}}
\def\mn@eprint@dblp#1{\href {http://dblp.uni-trier.de/rec/bibtex/#1.xml}
  {dblp:#1}}
\def\mn@eprint@#1:#2:#3:#4\@nil{\def\@tempa {#1}\def\@tempb {#2}\def\@tempc
  {#3}\ifx \@tempc \@empty \let \@tempc \@tempb \let \@tempb \@tempa \fi \ifx
  \@tempb \@empty \def\@tempb {arXiv}\fi \@ifundefined
  {mn@eprint@\@tempb}{\@tempb:\@tempc}{\expandafter \expandafter \csname
  mn@eprint@\@tempb\endcsname \expandafter{\@tempc}}}

\bibitem[\protect\citeauthoryear{{Bandura} et~al.,}{{Bandura}
  et~al.}{2014}]{2014SPIE.9145E..22B}
{Bandura} K.,  et~al., 2014, in Ground-based and Airborne Telescopes V. p.
  914522 (\mn@eprint {arXiv} {1406.2288}), \mn@doi{10.1117/12.2054950}

\bibitem[\protect\citeauthoryear{{Battye} et~al.,}{{Battye}
  et~al.}{2012}]{2012arXiv1209.1041B}
{Battye} R.~A.,  et~al., 2012, ArXiv, \href
  {http://adsabs.harvard.edu/abs/2012arXiv1209.1041B} {1209.1041}

\bibitem[\protect\citeauthoryear{{Battye} et~al.,}{{Battye}
  et~al.}{2016}]{2016arXiv161006826B}
{Battye} R.,  et~al., 2016, ArXiv, \href
  {http://adsabs.harvard.edu/abs/2016arXiv161006826B} {1610.06826}

\bibitem[\protect\citeauthoryear{{Bigiel}, {Leroy}, {Walter}, {Blitz},
  {Brinks}, {de Blok}  \& {Madore}}{{Bigiel}
  et~al.}{2010}]{2010AJ....140.1194B}
{Bigiel} F.,  {Leroy} A.,  {Walter} F.,  {Blitz} L.,  {Brinks} E.,  {de Blok}
  W.~J.~G.,   {Madore} B.,  2010, \mn@doi [\aj] {10.1088/0004-6256/140/5/1194},
  \href {https://ui.adsabs.harvard.edu/abs/2010AJ....140.1194B} {140, 1194}

\bibitem[\protect\citeauthoryear{{Bigot-Sazy} et~al.,}{{Bigot-Sazy}
  et~al.}{2015}]{2015MNRAS.454.3240B}
{Bigot-Sazy} M.~A.,  et~al., 2015, \mn@doi [\mnras] {10.1093/mnras/stv2153},
  \href {https://ui.adsabs.harvard.edu/abs/2015MNRAS.454.3240B} {454, 3240}

\bibitem[\protect\citeauthoryear{{Bigot-Sazy} et~al.,}{{Bigot-Sazy}
  et~al.}{2016}]{2016ASPC..502...41B}
{Bigot-Sazy} M.~A.,  et~al., 2016, in {Qain} L.,  {Li} D.,  eds,  Astronomical
  Society of the Pacific Conference Series Vol. 502, Frontiers in Radio
  Astronomy and FAST Early Sciences Symposium 2015. p.~41 (\mn@eprint {arXiv}
  {1511.03006})

\bibitem[\protect\citeauthoryear{{Chang}, {Pen}, {Peterson}  \&
  {McDonald}}{{Chang} et~al.}{2008}]{2008PhRvL.100i1303C}
{Chang} T.-C.,  {Pen} U.-L.,  {Peterson} J.~B.,   {McDonald} P.,  2008, \mn@doi
  [Physical Review Letters] {10.1103/PhysRevLett.100.091303}, \href
  {http://adsabs.harvard.edu/abs/2008PhRvL.100i1303C} {100, 091303}

\bibitem[\protect\citeauthoryear{{Chen}}{{Chen}}{2012}]{2012IJMPS..12..256C}
{Chen} X.,  2012, in International Journal of Modern Physics Conference Series.
  pp 256--263 (\mn@eprint {arXiv} {1212.6278}),
  \mn@doi{10.1142/S2010194512006459}

\bibitem[\protect\citeauthoryear{Chen, Battye, Costa, Dickinson  \&
  Harper}{Chen et~al.}{2020}]{Chen:2019jms}
Chen T.,  Battye R.~A.,  Costa A.~A.,  Dickinson C.,   Harper S.~E.,  2020,
  \mn@doi [Mon. Not. Roy. Astron. Soc.] {10.1093/mnras/stz3307}, 491, 4254

\bibitem[\protect\citeauthoryear{{Cunnington}, {Irfan}, {Carucci}, {Pourtsidou}
   \& {Bobin}}{{Cunnington} et~al.}{2020}]{2020arXiv201002907C}
{Cunnington} S.,  {Irfan} M.~O.,  {Carucci} I.~P.,  {Pourtsidou} A.,   {Bobin}
  J.,  2020, arXiv e-prints, \href
  {https://ui.adsabs.harvard.edu/abs/2020arXiv201002907C} {p. arXiv:2010.02907}

\bibitem[\protect\citeauthoryear{{Furlanetto} et~al.,}{{Furlanetto}
  et~al.}{2019}]{2019arXiv190306212F}
{Furlanetto} S.,  et~al., 2019, arXiv e-prints, \href
  {https://ui.adsabs.harvard.edu/abs/2019arXiv190306212F} {p. arXiv:1903.06212}

\bibitem[\protect\citeauthoryear{{Giovanelli} et~al.,}{{Giovanelli}
  et~al.}{2005}]{2005AJ....130.2598G}
{Giovanelli} R.,  et~al., 2005, \mn@doi [\aj] {10.1086/497431}, \href
  {http://adsabs.harvard.edu/abs/2005AJ....130.2598G} {130, 2598}

\bibitem[\protect\citeauthoryear{{Giovanelli} et~al.,}{{Giovanelli}
  et~al.}{2007}]{2007AJ....133.2569G}
{Giovanelli} R.,  et~al., 2007, \mn@doi [\aj] {10.1086/516635}, \href
  {http://adsabs.harvard.edu/abs/2007AJ....133.2569G} {133, 2569}

\bibitem[\protect\citeauthoryear{{Harper}, {Dickinson}, {Battye},
  {Roychowdhury}, {Browne}, {Ma}, {Olivari}  \& {Chen}}{{Harper}
  et~al.}{2018}]{2018MNRAS.478.2416H}
{Harper} S.~E.,  {Dickinson} C.,  {Battye} R.~A.,  {Roychowdhury} S.,  {Browne}
  I.~W.~A.,  {Ma} Y.~Z.,  {Olivari} L.~C.,   {Chen} T.,  2018, \mn@doi [\mnras]
  {10.1093/mnras/sty1238}, \href
  {https://ui.adsabs.harvard.edu/abs/2018MNRAS.478.2416H} {478, 2416}

\bibitem[\protect\citeauthoryear{Hu, Wang, Wu, Wang, Zhang  \& Chen}{Hu
  et~al.}{2020}]{Hu:2019okh}
Hu W.,  Wang X.,  Wu F.,  Wang Y.,  Zhang P.,   Chen X.,  2020, \mn@doi [Mon.
  Not. Roy. Astron. Soc.] {10.1093/mnras/staa650}, 493, 5854

\bibitem[\protect\citeauthoryear{Huang, Wu  \& Chen}{Huang
  et~al.}{2019}]{Huang:2018ral}
Huang Q.,  Wu F.,   Chen X.,  2019, \mn@doi [Sci. China Phys. Mech. Astron.]
  {10.1007/s11433-018-9333-1}, 62, 989511

\bibitem[\protect\citeauthoryear{{Janssen} et~al.,}{{Janssen}
  et~al.}{1996}]{1996astro.ph..2009J}
{Janssen} M.~A.,  et~al., 1996, arXiv e-prints, \href
  {https://ui.adsabs.harvard.edu/abs/1996astro.ph..2009J} {pp
  astro--ph/9602009}

\bibitem[\protect\citeauthoryear{{Jarvis} et~al.,}{{Jarvis}
  et~al.}{2014}]{2014arXiv1401.4018J}
{Jarvis} M.~J.,  et~al., 2014, preprint, \href
  {http://adsabs.harvard.edu/abs/2014arXiv1401.4018J} {} (\mn@eprint {arXiv}
  {1401.4018})

\bibitem[\protect\citeauthoryear{{Jiang} et~al.,}{{Jiang}
  et~al.}{2020}]{2020RAA....20...64J}
{Jiang} P.,  et~al., 2020, \mn@doi [Research in Astronomy and Astrophysics]
  {10.1088/1674-4527/20/5/64}, \href
  {https://ui.adsabs.harvard.edu/abs/2020RAA....20...64J} {20, 064}

\bibitem[\protect\citeauthoryear{{Jones}, {Papastergis}, {Haynes}  \&
  {Giovanelli}}{{Jones} et~al.}{2016}]{2016MNRAS.457.4393J}
{Jones} M.~G.,  {Papastergis} E.,  {Haynes} M.~P.,   {Giovanelli} R.,  2016,
  \mn@doi [\mnras] {10.1093/mnras/stw263}, \href
  {https://ui.adsabs.harvard.edu/abs/2016MNRAS.457.4393J} {457, 4393}

\bibitem[\protect\citeauthoryear{{Keih{\"a}nen}, {Kurki-Suonio}, {Poutanen},
  {Maino}  \& {Burigana}}{{Keih{\"a}nen} et~al.}{2004}]{2004A&A...428..287K}
{Keih{\"a}nen} E.,  {Kurki-Suonio} H.,  {Poutanen} T.,  {Maino} D.,
  {Burigana} C.,  2004, \mn@doi [\aap] {10.1051/0004-6361:200400060}, \href
  {https://ui.adsabs.harvard.edu/abs/2004A&A...428..287K} {428, 287}

\bibitem[\protect\citeauthoryear{{Kennicutt}}{{Kennicutt}}{1998}]{1998ApJ...498..541K}
{Kennicutt} Jr. R.~C.,  1998, \mn@doi [\apj] {10.1086/305588}, \href
  {http://adsabs.harvard.edu/abs/1998ApJ...498..541K} {498, 541}

\bibitem[\protect\citeauthoryear{{Krumholz}}{{Krumholz}}{2012}]{2012ApJ...759....9K}
{Krumholz} M.~R.,  2012, \mn@doi [\apj] {10.1088/0004-637X/759/1/9}, \href
  {https://ui.adsabs.harvard.edu/abs/2012ApJ...759....9K} {759, 9}

\bibitem[\protect\citeauthoryear{{Kurki-Suonio}, {Keih{\"a}nen}, {Keskitalo},
  {Poutanen}, {Sirvi{\"o}}, {Maino}  \& {Burigana}}{{Kurki-Suonio}
  et~al.}{2009}]{2009A&A...506.1511K}
{Kurki-Suonio} H.,  {Keih{\"a}nen} E.,  {Keskitalo} R.,  {Poutanen} T.,
  {Sirvi{\"o}} A.~S.,  {Maino} D.,   {Burigana} C.,  2009, \mn@doi [\aap]
  {10.1051/0004-6361/200912361}, \href
  {https://ui.adsabs.harvard.edu/abs/2009A&A...506.1511K} {506, 1511}

\bibitem[\protect\citeauthoryear{{Lelli}, {Verheijen}  \& {Fraternali}}{{Lelli}
  et~al.}{2014}]{2014A&A...566A..71L}
{Lelli} F.,  {Verheijen} M.,   {Fraternali} F.,  2014, \mn@doi [\aap]
  {10.1051/0004-6361/201322657}, \href
  {https://ui.adsabs.harvard.edu/abs/2014A&A...566A..71L} {566, A71}

\bibitem[\protect\citeauthoryear{{Li} et~al.,}{{Li}
  et~al.}{2018}]{2018IMMag..19..112L}
{Li} D.,  et~al., 2018, \mn@doi [IEEE Microwave Magazine]
  {10.1109/MMM.2018.2802178}, \href
  {https://ui.adsabs.harvard.edu/abs/2018IMMag..19..112L} {19, 112}

\bibitem[\protect\citeauthoryear{{Li}, {Santos}, {Grainge}, {Harper}  \&
  {Wang}}{{Li} et~al.}{2020a}]{2020arXiv200701767L}
{Li} Y.,  {Santos} M.~G.,  {Grainge} K.,  {Harper} S.,   {Wang} J.,  2020a,
  arXiv e-prints, \href {https://ui.adsabs.harvard.edu/abs/2020arXiv200701767L}
  {p. arXiv:2007.01767}

\bibitem[\protect\citeauthoryear{Li et~al.}{Li et~al.}{2020b}]{Li:2020ast}
Li J.,  et~al., 2020b, \mn@doi [Sci. China Phys. Mech. Astron.]
  {10.1007/s11433-020-1594-8}, 63, 129862

\bibitem[\protect\citeauthoryear{{Liu} \& {Shaw}}{{Liu} \&
  {Shaw}}{2020}]{2020PASP..132f2001L}
{Liu} A.,  {Shaw} J.~R.,  2020, \mn@doi [\pasp] {10.1088/1538-3873/ab5bfd},
  \href {https://ui.adsabs.harvard.edu/abs/2020PASP..132f2001L} {132, 062001}

\bibitem[\protect\citeauthoryear{{Maino}, {Burigana}, {G{\'o}rski}, {Mandolesi}
   \& {Bersanelli}}{{Maino} et~al.}{2002}]{2002A&A...387..356M}
{Maino} D.,  {Burigana} C.,  {G{\'o}rski} K.~M.,  {Mandolesi} N.,
  {Bersanelli} M.,  2002, \mn@doi [\aap] {10.1051/0004-6361:20020242}, \href
  {https://ui.adsabs.harvard.edu/abs/2002A&A...387..356M} {387, 356}

\bibitem[\protect\citeauthoryear{{Matshawule}, {Spinelli}, {Santos}  \&
  {Ngobese}}{{Matshawule} et~al.}{2020}]{2020arXiv201110815M}
{Matshawule} S.~D.,  {Spinelli} M.,  {Santos} M.~G.,   {Ngobese} S.,  2020,
  arXiv e-prints, \href {https://ui.adsabs.harvard.edu/abs/2020arXiv201110815M}
  {p. arXiv:2011.10815}

\bibitem[\protect\citeauthoryear{{Meyer} et~al.,}{{Meyer}
  et~al.}{2004}]{2004MNRAS.350.1195M}
{Meyer} M.~J.,  et~al., 2004, \mn@doi [\mnras]
  {10.1111/j.1365-2966.2004.07710.x}, \href
  {http://adsabs.harvard.edu/abs/2004MNRAS.350.1195M} {350, 1195}

\bibitem[\protect\citeauthoryear{{Milotti}}{{Milotti}}{2002}]{2002physics...4033M}
{Milotti} E.,  2002, arXiv e-prints, \href
  {https://ui.adsabs.harvard.edu/abs/2002physics...4033M} {p. physics/0204033}

\bibitem[\protect\citeauthoryear{{Morales} \& {Wyithe}}{{Morales} \&
  {Wyithe}}{2010}]{2010ARA&A..48..127M}
{Morales} M.~F.,  {Wyithe} J. S.~B.,  2010, \mn@doi [\araa]
  {10.1146/annurev-astro-081309-130936}, \href
  {https://ui.adsabs.harvard.edu/abs/2010ARA&A..48..127M} {48, 127}

\bibitem[\protect\citeauthoryear{{Mundell} \& {Shone}}{{Mundell} \&
  {Shone}}{1999}]{1999MNRAS.304..475M}
{Mundell} C.~G.,  {Shone} D.~L.,  1999, \mn@doi [\mnras]
  {10.1046/j.1365-8711.1999.02330.x}, \href
  {https://ui.adsabs.harvard.edu/abs/1999MNRAS.304..475M} {304, 475}

\bibitem[\protect\citeauthoryear{{Nan} et~al.,}{{Nan}
  et~al.}{2011}]{2011IJMPD..20..989N}
{Nan} R.,  et~al., 2011, \mn@doi [International Journal of Modern Physics D]
  {10.1142/S0218271811019335}, \href
  {http://adsabs.harvard.edu/abs/2011IJMPD..20..989N} {20, 989}

\bibitem[\protect\citeauthoryear{{Newburgh} et~al.,}{{Newburgh}
  et~al.}{2014}]{2014SPIE.9145E..4VN}
{Newburgh} L.~B.,  et~al., 2014, in {Stepp} L.~M.,  {Gilmozzi} R.,   {Hall}
  H.~J.,  eds,  Society of Photo-Optical Instrumentation Engineers (SPIE)
  Conference Series Vol. 9145, Ground-based and Airborne Telescopes V. p.
  91454V (\mn@eprint {arXiv} {1406.2267}), \mn@doi{10.1117/12.2056962}

\bibitem[\protect\citeauthoryear{{Newburgh} et~al.,}{{Newburgh}
  et~al.}{2016}]{2016SPIE.9906E..5XN}
{Newburgh} L.~B.,  et~al., 2016, in Ground-based and Airborne Telescopes VI. p.
  99065X (\mn@eprint {arXiv} {1607.02059}), \mn@doi{10.1117/12.2234286}

\bibitem[\protect\citeauthoryear{{Oosterloo} et~al.,}{{Oosterloo}
  et~al.}{2010}]{2010MNRAS.409..500O}
{Oosterloo} T.,  et~al., 2010, \mn@doi [\mnras]
  {10.1111/j.1365-2966.2010.17351.x}, \href
  {https://ui.adsabs.harvard.edu/abs/2010MNRAS.409..500O} {409, 500}

\bibitem[\protect\citeauthoryear{{Planck Collaboration} et~al.,}{{Planck
  Collaboration} et~al.}{2018}]{2018arXiv180706209P}
{Planck Collaboration} et~al., 2018, arXiv e-prints, \href
  {https://ui.adsabs.harvard.edu/abs/2018arXiv180706209P} {p. arXiv:1807.06209}

\bibitem[\protect\citeauthoryear{{Pritchard} \& {Loeb}}{{Pritchard} \&
  {Loeb}}{2012}]{2012RPPh...75h6901P}
{Pritchard} J.~R.,  {Loeb} A.,  2012, \mn@doi [Reports on Progress in Physics]
  {10.1088/0034-4885/75/8/086901}, \href
  {https://ui.adsabs.harvard.edu/abs/2012RPPh...75h6901P} {75, 086901}

\bibitem[\protect\citeauthoryear{{Saintonge}}{{Saintonge}}{2007}]{2007AJ....133.2087S}
{Saintonge} A.,  2007, \mn@doi [\aj] {10.1086/513515}, \href
  {http://adsabs.harvard.edu/abs/2007AJ....133.2087S} {133, 2087}

\bibitem[\protect\citeauthoryear{{Santos} et~al.,}{{Santos}
  et~al.}{2015}]{2015aska.confE..19S}
{Santos} M.,  et~al., 2015, in Advancing Astrophysics with the Square Kilometre
  Array (AASKA14). p.~19 (\mn@eprint {arXiv} {1501.03989})

\bibitem[\protect\citeauthoryear{{Santos} et~al.,}{{Santos}
  et~al.}{2017}]{2017arXiv170906099S}
{Santos} M.~G.,  et~al., 2017, arXiv e-prints, \href
  {https://ui.adsabs.harvard.edu/abs/2017arXiv170906099S} {p. arXiv:1709.06099}

\bibitem[\protect\citeauthoryear{{Schmidt}}{{Schmidt}}{1959}]{1959ApJ...129..243S}
{Schmidt} M.,  1959, \mn@doi [\apj] {10.1086/146614}, \href
  {http://adsabs.harvard.edu/abs/1959ApJ...129..243S} {129, 243}

\bibitem[\protect\citeauthoryear{{Seiffert}, {Mennella}, {Burigana}, {Mand
  olesi}, {Bersanelli}, {Meinhold}  \& {Lubin}}{{Seiffert}
  et~al.}{2002}]{2002A&A...391.1185S}
{Seiffert} M.,  {Mennella} A.,  {Burigana} C.,  {Mand olesi} N.,  {Bersanelli}
  M.,  {Meinhold} P.,   {Lubin} P.,  2002, \mn@doi [\aap]
  {10.1051/0004-6361:20020880}, \href
  {https://ui.adsabs.harvard.edu/abs/2002A&A...391.1185S} {391, 1185}

\bibitem[\protect\citeauthoryear{{Sutton} et~al.,}{{Sutton}
  et~al.}{2010}]{2010MNRAS.407.1387S}
{Sutton} D.,  et~al., 2010, \mn@doi [\mnras]
  {10.1111/j.1365-2966.2010.16954.x}, \href
  {https://ui.adsabs.harvard.edu/abs/2010MNRAS.407.1387S} {407, 1387}

\bibitem[\protect\citeauthoryear{Van~Rossum \& Drake~Jr}{Van~Rossum \&
  Drake~Jr}{1995}]{van1995python}
Van~Rossum G.,  Drake~Jr F.~L.,  1995, Python reference manual.
Centrum voor Wiskunde en Informatica Amsterdam

\bibitem[\protect\citeauthoryear{Virtanen et~al.,}{Virtanen
  et~al.}{2020}]{2020SciPy-NMeth}
Virtanen P.,  et~al., 2020, \mn@doi [Nature Methods]
  {10.1038/s41592-019-0686-2}, \href {https://rdcu.be/b08Wh} {17, 261}

\bibitem[\protect\citeauthoryear{Wu et~al.,}{Wu et~al.}{2020}]{wu2020tianlai}
Wu F.,  et~al., 2020, The Tianlai Dish Pathfinder Array: design, operation and
  performance of a prototype transit radio interferometer (\mn@eprint {arXiv}
  {2011.05946})

\bibitem[\protect\citeauthoryear{{Xu}, {Wang}  \& {Chen}}{{Xu}
  et~al.}{2015}]{2015ApJ...798...40X}
{Xu} Y.,  {Wang} X.,   {Chen} X.,  2015, \mn@doi [\apj]
  {10.1088/0004-637X/798/1/40}, \href
  {http://adsabs.harvard.edu/abs/2015ApJ...798...40X} {798, 40}

\bibitem[\protect\citeauthoryear{{Zheng} et~al.,}{{Zheng}
  et~al.}{2017}]{2017MNRAS.464.3486Z}
{Zheng} H.,  et~al., 2017, \mn@doi [\mnras] {10.1093/mnras/stw2525}, \href
  {https://ui.adsabs.harvard.edu/abs/2017MNRAS.464.3486Z} {464, 3486}

\bibitem[\protect\citeauthoryear{{Zwaan} et~al.,}{{Zwaan}
  et~al.}{2004}]{2004MNRAS.350.1210Z}
{Zwaan} M.~A.,  et~al., 2004, \mn@doi [\mnras]
  {10.1111/j.1365-2966.2004.07782.x}, \href
  {http://adsabs.harvard.edu/abs/2004MNRAS.350.1210Z} {350, 1210}

\bibitem[\protect\citeauthoryear{{de Oliveira-Costa}, {Tegmark}, {Gaensler},
  {Jonas}, {Landecker}  \& {Reich}}{{de Oliveira-Costa}
  et~al.}{2008}]{2008MNRAS.388..247D}
{de Oliveira-Costa} A.,  {Tegmark} M.,  {Gaensler} B.~M.,  {Jonas} J.,
  {Landecker} T.~L.,   {Reich} P.,  2008, \mn@doi [\mnras]
  {10.1111/j.1365-2966.2008.13376.x}, \href
  {https://ui.adsabs.harvard.edu/abs/2008MNRAS.388..247D} {388, 247}

\makeatother
\end{thebibliography}

\bsp	
\label{lastpage}
\end{document}